\documentstyle[eqsecnum,aps,epsfig]{revtex}

\begin{document} \draft \preprint{HEP/123-qed}
\title{Security of Quantum Key Distribution with Entangled Photons Against Individual Attacks}
\author{Edo Waks, Assaf Zeevi, Yoshihisa Yamamoto\footnote{Also at NTT Basic Research Laboratories, Atsugi, Kanagawa, Japan.}}
\address{Quantum Entanglement Project, ICORP, JST, E.L. Ginzton Laboratory, Stanford University \\ Stanford, CA 94305}
\date{\today}
\maketitle

\newcommand{\avg}[1]{\langle #1 \rangle}
\def\Re{\mbox{Re}}
\def\Im{\mbox{Im}}
\newcommand{\TR}[1]{\mbox{Tr}\left\{ #1 \right\}}
\def\<#1|{\langle#1|}
\def\|#1>{|#1\rangle}
\def\Trans{\alpha_L}
\def\T2{\alpha_{L/2}}
\def\TM{\alpha_{N}}
\def\nbar{\bar{n}}
\def\adag{\hat{a}^{\dagger}}
\def\bdag{\hat{b}^{\dagger}}
\newcommand{\n}{a^\dagger a}
\newcommand{\tr}{\mbox{Tr}}
\def\bk<#1|#2>{\left\langle\vphantom{#1|#2}#1\right|%
\left.\vphantom{#1|#2}#2\right\rangle}
\def\kb|#1><#2|{\left|\vphantom{#1|#2}#1\right\rangle%
\left\langle\vphantom{#1|#2}#2\right|}
\def\aux|#1><#2|{\left|\vphantom{#1|#2}#1\right\rangle_{a}%
\left\langle\vphantom{#1|#2}#2\right|}

\begin{abstract}

Security of the Ekert protocol is proven against individual
attacks where an eavesdropper is allowed to share any density
matrix with the two communicating parties.  The density matrix
spans all of the photon number states of both receivers, as well
as a probe state of arbitrary dimensionality belonging to the
eavesdropper.  Using this general eavesdropping strategy, we show
that the Shannon information on the final key, after error
correction and privacy amplification, can be made exponentially
small. This is done by finding a bound on the eavesdropper's
average collision probability.  We find that the average collision
probability for the Ekert protocol is the same as that of the BB84
protocol for single photons, indicating that there is no analog in
the Ekert protocol to photon splitting attacks. We then compare
the communication rate of both protocols as a function of
distance, and show that the Ekert protocol has potential for much
longer communication distances, up to $170$km, in the presence of
realistic detector dark counts and channel loss. Finally, we
propose a slightly more complicated scheme based on entanglement
swapping that can lead to even longer distances of communication.
The limiting factor in this new scheme is the fiber loss, which
imposes very slow communication rates at longer distances.
\end{abstract}

\section{Introduction}

The field of quantum information theory has brought the potential
to accomplish feats considered impossible by purely classical
methods. One of these is the ability to transmit an
unconditionally secure message between two parties, known as
quantum cryptography.  The first full protocol for quantum
cryptography was proposed by Bennett and Brassard using four
different states of a quantum system~\cite{BennettBrassard84}, and
has since been known as BB84. Following the discovery of BB84,
other protocols such as the two-state and six-state schemes have
been proposed~\cite{Bennett92,HuttnerImoto95}.  The security of
all of these protocols relies on the impossibility of an
eavesdropper to measure the wavefunction of a quantum system
without imposing a backaction on the state. This backaction will
usually result in a measurable increase in errors across the
communication channel.

In 1991 it was proposed by Ekert that quantum key distribution
could also be implemented using non-local correlations between
quantum systems~\cite{Ekert91}.  Ekert pointed out that two
correlated quantum systems cannot lead to violations of Bell's
inequality if they are also correlated to a local variable which
an eavesdropper can observe in order to gain knowledge of the
measurement results.  A test of Bell's inequality could then
provide a statement of security against eavesdropping. It was
later discovered that Bell's inequality is not necessary for
security of Ekert's protocol~\cite{BennettBrassard92}.  An
eavesdropper cannot obtain knowledge from a correlated quantum
state without inducing errors, just as in the other protocols.

The experimental effort to perform quantum key distribution began
soon after the theory was established.  Several groups have
reported implementations of BB84 and other single photon
schemes~\cite{ButtlerHughes00,Townsend98,MarandTownsend95,RigbordyGautier00,BourennaneGibson99}.
Long distance violations of Bell's inequality have been
demonstrated in~\cite{TittelBrendel99}, and recently several
proof-of-principle experiments using entangled photons have also
been
performed~\cite{JenneweinSimon00,TittelBrendel00,NaikPeterson00}.
For experimental quantum cryptography it is insufficient to show
that tampering with the quantum channel will always result in some
error.  Practical systems have a baseline error rate which cannot
be distinguished from tampering. These errors can be handled by
public discussion through two additional steps, error correction
and privacy amplification.  The error correction step serves the
dual purpose of correcting all erroneously received bits and
giving an estimate of the error rate. Privacy amplification is
then used to distill a shorter key which can be made as secure as
desired.  The length of this key depends on the amount of
information which may have been leaked, and this should ideally be
determined from the measured error rate and the laws of quantum
mechanics.

The security of quantum key distribution against any attack
allowed by the laws of quantum mechanics is a complex subject.
Earlier work showed security for several restricted types of
attacks~\cite{Ekert94,HuttnerEkert94}.  Later security was proved
for the most general individual attacks in
BB84~\cite{Fuchs97,Slutsky98,Lutkenhaus99}, and these proofs were
extended to practical photon sources in~\cite{Lutkenhaus00}. In an
individual attack the eavesdropper is restricted to measuring each
quantum transmission independently, but is allowed to use any
measurement which is not forbidden by quantum mechanics.  A more
general attack allows collective measurements which make use of
the correlations introduced during error correction and privacy
amplification by exchange of block parities. This information can
be used to refine an eavesdropper's quantum measurement.  Security
against these more general attacks has been shown
in~\cite{BihamMor97}. The most general type of attack is known as
a joint attack where the eavesdropper treats the entire quantum
transmission as one system which she entangles with a probe of
very large dimensionality. There are currently several proofs of
security against this most general
scenario~\cite{Mayers96,LoChau99,BihamBoyer00,ShorPreskill00}.
Security against collective and joint attacks is an important
milestone in quantum information theory, however the current
proofs of security are difficult to apply to realistic systems.
Furthermore, it seems practically impossible to implement such
attacks using currently available technology. On the other hand
practical systems can be rendered completely insecure even by very
simple measurements~\cite{BrassardLutkenhaus00}. Thus, restricting
Eve to only individual attacks is usually deemed reasonable for
practical purposes.

The security of the Ekert protocol has not been studied as closely
as that of BB84.  Unlike BB84, in the Ekert protocol the photon
source is located somewhere between the sender and receiver and
provides both with photons.  Since the two communicating parties
no longer have control over the source, the eavesdropper is not
restricted to making measurements.  She can take on the more
aggressive strategy of blocking out the source and providing the
two parties with her own photons. For example, she can provide
each receiver with a photon in a known quantum state. This is the
two-photon analog to the "intercept and re-send" attack in BB84,
and will always result in a $25\%$ error rate.  A more general
eavesdropping attack allows Eve to send one photon to each
receiver, while maintaining a probe system that can be measured
later on. This model was used in~\cite{BennettBrassard92} to show
that if the eavesdropper insisted on not causing any errors, then
she could not obtain any information about the measurements at the
two receivers.  However, as previously mentioned, this is not
sufficient for practical quantum key distribution. We need to know
the explicit relationship between the error rate and amount of
information leaked, which has yet to be given. Furthermore, an
eavesdropper could send more than one photon to either receiver,
and a proof of security must take this into account.

In this paper we prove the security of the Ekert protocol against
the most general types of individual attacks.  In such an attack
the eavesdropper is allowed to share any density matrix
$\rho_{abe}$ with the two receivers.  This density matrix
describes the state of the signal sent to each receiver, and a
probe state of arbitrary dimensions which the eavesdropper will
use to infer information.  Security is proved by upper bounding
Eve's mutual information on the final key, as well as explicitly
deriving an equation for the length of this key after privacy
amplification. We make no idealized assumptions about the source,
such as that it emits only one pair of photons per clock cycle.
Thus, our results can be directly applied to practical key
distribution schemes. One interesting aspect of our result is that
the relationship between error rate and leaked information for the
Ekert protocol with any source is the same as that of BB84 with an
ideal single photon device. This indicates that, at least against
individual attacks, there is no analog in the Ekert protocol to
the powerful photon splitting attacks which severely jeopardize
the security of BB84. Using the derived expression for the length
of the final key, we calculate the communication rate for the
Ekert protocol with ideal entangled photon sources, as well as
realistic sources based on parametric down-conversion. We compare
these rates with the communication rate for BB84 using poissonian,
sub-poissonian, and ideal photon sources.  It should be stated
that some of the derivations in this paper are involved, but the
results themselves are extremely easy to use, involving simple
functions of experimentally measurable quantities.  A concise
review of the important equations is given in
Section~\ref{discussion}.

In Section~\ref{protocol} we review the general theory behind
quantum key distribution.  We restate some important information
theoretic results on error correction and privacy amplification.
We then derive a method for handling the side information leaked
during error correction.  This method allows us to account for the
effect of error correction on the length of the final key.  We
also re-derive rates for BB84 using both poissonian and
sub-poissonian light sources. These rates will later be used to
make a comparison to the Ekert protocol.  In Section~\ref{rates}
we derive a proof of security for the Ekert protocol, and use it
to calculate expected communication rates under practical
experimental conditions. Finally, in Section~\ref{swap} we
investigate an experimental configuration based on entanglement
swapping which is less sensitive to channel loss and detector dark
counts. With some technological improvement this configuration may
be useful for long distance quantum key distribution.

\section{Preliminaries} \label{protocol}

In this section we provide a concise review of important concepts
in quantum key distribution (QKD).  We also derive some
preliminary results which we will use in the upcoming sections.
The standard participants in QKD are Alice, Bob, and Eve. Alice
would like to exchange a secret key with Bob, which can later be
used to encode the actual message. To do this she uses both a
quantum channel and a public channel. The enemy, Eve, can listen
in on the public channel, but is assumed incapable of altering the
messages being exchanged. Eve is also allowed to make any
measurements she can on the quantum channel.

The secret key is formed in three steps.  The first is the raw
quantum transmission, which uses both the quantum and public
channel simultaneously.  The next two steps, error correction and
privacy amplification, make use of only the public channel. After
privacy amplification Alice and Bob each posses a copy of the
secret key, about which Eve knows only a negligibly small amount
of information.

\subsection{Quantum transmission} \label{QuanTrans}

  \begin{figure}
    \centering
    \epsfxsize = 5in
    \epsfbox{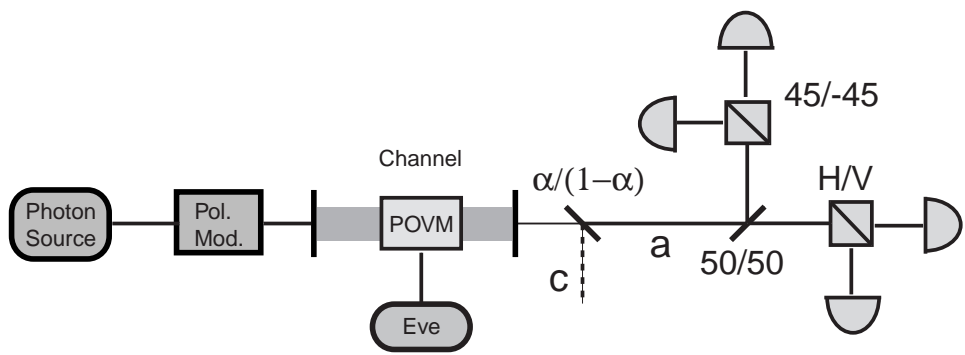}
    \caption{Schematic of experimental setup for BB84.}\label{BB84}
  \end{figure}

In the BB84 protocol Alice sends Bob a sequence of individual
quantum bits (qubits) which randomly encode binary $0$ or $1$. The
qubits are assumed to be photons with the information encoded in
the polarization, but other physical implementations can usually
be treated in an analogous way. Alice uses two different
non-orthogonal bases to encode her information. For example, half
of the time she may encode her information along the $x$-$y$ axes,
and the other half she uses the $45^o$ rotated axes which we will
refer to as $u$-$v$. Bob randomly chooses one of the two bases and
makes a projective measurement.   After all of the quantum bits
have been exchanged, Alice and Bob will publicly disclose the
bases they used, but not the measurement results. They agree to
keep only the bits in which both parties used the same basis.
These bits form the sifted key.

Figure~\ref{BB84} shows a schematic of the BB84 protocol.  Bob is
looking for photons in a spatial mode which we will label as mode
$a$.  The photons in this mode are randomly partitioned by a
$50/50$ beamsplitter and sent to one of the two polarizing
beamsplitter.  This technique, known as passive modulation, is an
easy way to randomly modulate Bob's measurement basis.  Passive
modulation was chosen because it is easier to implement in
practical systems, and also because it simplifies the proof of
security~\cite{Lutkenhaus99}. In order to model the loss in the
quantum channel, detectors, and optics,  we introduce an
additional beamsplitter with transmission $\alpha$ and a loss mode
$c$.  We can set the value of $\alpha$ to the total loss of the
system and assume that the channel, detectors, and optics are
lossless.  The advantage of this approach is that it allows us to
easily treat the effect of losses on the quantum state of the
incoming photon. All we need to do is apply a beamsplitter
transformation onto the photon and trace out over mode $c$ to get
its final state.  This is not as important in the BB84 protocol
since the effect of the losses is rather obvious. Either a photon
is detected or it is not. However, when we analyze the Ekert
protocol this model will prove to be extremely helpful.

Figure~\ref{BB84} also shows the role of Eve.  Eve tries to
measure the state of each transmission that has been sent into the
quantum channel. The most general type of individual measurement
Eve can perform is a Positive Operator Value Measurement
(POVM)\cite{Helstrom76}.   In this type of measurement Eve
entangles a quantum mechanical probe with each photon through a
unitary evolution. The probe is stored coherently until all
information from public discussion is revealed.  Eve then uses all
publicly disclosed information to make the best measurement on her
probe. Her only restriction is that she measures each probe
independently in compliance with the assumption of individual
attacks.  Any POVM can be characterized by a complete set of
positive maps $A_k$ which satisfy the condition
  \begin{equation}
    \sum_k A_k A_k^{\dagger} = I.
  \end{equation}
If Alice sends the signal in a quantum state described by the
density matrix $\rho$, the measurement backaction on the state is
described by
  \begin{equation}
    \tilde{\rho} = \sum_k A_k \rho A_k^{\dagger},
  \end{equation}
where $\tilde{\rho}$ is the quantum state after the measurement.
A pure state can evolve into a mixed state through this type of
intervention. The probability that Eve will measure her probe in
the state $k$ is given by
  \begin{equation}
    p(k) = \TR{A_k \rho A_k^{\dagger}},
  \end{equation}
while the probability that Bob measures outcome $\psi$ is
  \begin{equation}
    p(\psi,k) = \TR{A_k \rho A_k^{\dagger} E_{\psi}},
  \end{equation}
where $E_{\psi}$ denotes the projection operator onto the state
$\|\psi>$.

Unlike BB84, in the Ekert protocol both Alice and Bob are
recipients of a signal.  The source of this signal is presumed to
be somewhere in between both parties.  In the ideal case each
party receives one of a pair of photons in a quantum mechanically
entangled state. For example, the two photons may be in the state
  \begin{equation}
    \|\psi> = \frac{1}{\sqrt{2}} \left( \|xx> + \|yy> \right),
  \end{equation}
which implies that if both receivers measure their photon in the
$x$-$y$ basis, their measurment results will be completely
correlated.  However, one can rewrite the above state in the
completely equivalent form
  \begin{equation}
    \|\psi> = \frac{1}{\sqrt{2}} \left( \|uu> + \|vv> \right).
  \end{equation}
Thus, if both receivers choose to measure in the $u$-$v$ instead
of the $x$-$y$ basis their measurement results will remain
correlated.  This suggests the following protocol for quantum
cryptography.  Each receiver measures their respective photon
randomly in either the $x$-$y$ or $u$-$v$ basis. Later they agree
to keep only the instances in which the measurement bases were the
same, forming the sifted key.

\begin{figure}
  \centering
  \epsfxsize = 5in
  \epsfbox{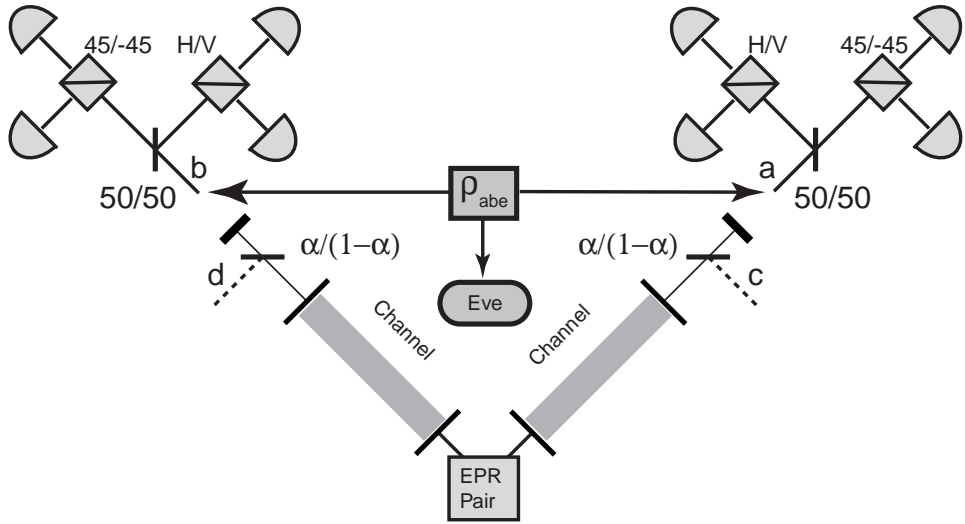}
  \caption{Schematic of experimental setup for Ekert protocol.}\label{EPR-QKD}
\end{figure}

Figure~\ref{EPR-QKD} shows a schematic of a key distribution
experiment based on the Ekert protocol.  In this experiment Alice
is assumed to monitor mode $a$, and Bob mode $b$.  Both parties
use passive modulation to switch their basis.  As before, we
insert an additional beamsplitter into each arm to account for the
losses and presume that the channel, optics, and detectors are all
lossless.

In the Ekert protocol, the role of the Eve is also modified. Since
she now has complete control over the source, Eve is not
restricted to making measurements on the photons.  In fact, she
can can completely block out the source and provide the two
receivers with any signal she chooses, as shown in
Figure~\ref{EPR-QKD}. For example, Eve can send both receivers a
photon polarized along the $x$ axis. If Alice and Bob measure in
the $x$-$y$ basis then she will know their measurement result. But
if they choose the $u$-$v$ basis, then their measurements are
completely random, resulting in a $50\%$ error rate.  This is the
equivalent version of the intercept and re-send strategy in BB84.
A more general attack strategy is to generate a pure state
consisting of one photon for Alice, one for Bob, and some probe
which Eve can later use to infer the measurement results. The most
general pure state of this type that Eve can generate is
  \begin{equation}
    \|\psi_{abe}> = \|xx>\|P_{xx}> + \|yy>\|P_{yy}> +
    \|xy>\|P_{xy}> + \|yx>\|P_{yx}>,
  \end{equation}
where $\|P_{xx}>$, $\|P_{yy}>$, $\|P_{xy}>$, and $\|P_{yx}>$ are
the states of her probe and are not assumed to be orthogonal or
normalized.  As with BB84, she can store her probe until all
public information is revealed, and use this to refine her
measurement. But even this is not completely general, because Eve
could send a mixed state instead of a pure state.  Furthermore,
she may decide to send more than one photon to either Alice or Bob
if this is advantageous.  The most general state that Eve can
generate is a density matrix $\rho_{abe}$, which spans the entire
Hilbert space of Eve's probe, as well as the entire photon number
manifold of Alice and Bob.  Once again it should be emphasized
that we are restricting Eve to individual attacks, which means
that she measures her probes independently, and that $\rho_{abe}$
is independently generated for each clock cycle.

It is important to point out one additional subtlety regarding
security of the Ekert protocol. In the above model we assume that
Eve is in fact sending photons, and not some other particle.  This
assumption may seem silly, but appears unavoidable, at least if
one insists on using passive modulation. The reason for this is
that both receivers take for granted that the optics they are
using perform the measurement which they intended.  They rely on
the polarizing beamsplitters to make a projective measurements in
the polarization basis, and the $50/50$ beamsplitter to randomly
partition the photons. Suppose that Eve has access to a mysterious
particle which is also capable of triggering a detection event in
the photon counters. However, assume that this particle has a spin
angular momentum of $3/2$. Furthermore, the particle interacts in
such a way that the $50/50$ beamsplitter reflects the $3/2$ and
$1/2$ spin states, but transmits $-3/2$ and $-1/2$.  The
polarizing beamsplitters instead reflect $3/2$ and $-3/2$, but
transmit $1/2$ and $-1/2$.  If Eve sends this strange unnamed
particle instead of a photon she has complete control over the
detection events at both receivers. Although it is highly unlikely
that such a particle exists, it is a very difficult claim to
prove.  It has been recently shown that this loophole can be
circumvented if Alice and Bob rapidly switch their measurement
basis~\cite{MayersYao98}.  However, this method requires very low
loss, which is extremely difficult to achieve with practical
systems.

\subsection{Error correction} \label{ErrCorr}

In any realistic communication system errors are bound to occur,
and some form of error correction is required.  In quantum
cryptography the errors typically arise from technological
imperfections in the optics and detectors, but can also come from
eavesdropping. In order to achieve noise free communication these
errors must be corrected, and this can be done through public
discussion.

Following the raw quantum transmission Alice, Bob, and Eve each
possess the strings $X$, $Y$, and $Z$ respectively.  In order to
correct the errors, Alice and Bob exchange an additional message
$U$ such that knowledge of string $Y$ and $U$ leave very little
uncertainty about string $X$.  One way to mathematically express
this is to use the Shannon entropy function~\cite{Cover}
  \begin{equation}
    H(X) = -\sum_x p(x) \log_2 p(x).
  \end{equation}
The conditional entropy function $H(X|Z=z)$ is defined as above
using the conditional probability distribution $p(x|Z=z)$.  The
average conditional entropy $H(X|Z)$ is simply defined as
  \begin{equation}
    H(X|Z) = \sum_z p(z) H(X|Z=z).
  \end{equation}
The message $U$ should provide Bob with enough information so that
$H(X|YU)~\approx~0$.  Since string $U$ is publicly disclosed Eve
may learn additional information as well, but good error
correction algorithm will reduce this information leakage to a
minimum. Unfortunately, given the error rate $e$, a lower bound
exists on the minimum number of bits in $U$. This limit, which is
a variant of the Shannon noiseless coding theorem can be stated as
  \begin{equation} \label{ShannonLim}
    \lim_{n\to\infty} \frac{\kappa}{n} \ge h \left( e
    \right),
  \end{equation}
where $n$ is the length of the string, $\kappa$ is the number of
bits in message $U$, and $h(e)$ is the conditional entropy of a
single bit over a binary symmetric channel which is given by
   \begin{equation} \label{ShannonEntropy}
     h \left( e \right) = -e\log e - \left( 1 - e \right) \log
     \left( 1 - e \right).
   \end{equation}
An error correction algorithm should ideally operate very close to
this limit.  At the same time the algorithm should be
computationally efficient or the execution time may become
prohibitively long.

Error correction algorithms can usually be divided into two
classes, unidirectional and bidirectional algorithms.  In a
unidirectional algorithm information flows only from Alice to Bob.
Alice provides Bob with an additional string $U$ which he then
uses to try to find his errors. Unidirectional algorithms that are
both computationally efficient and operate near the Shannon limit
are difficult to find~\cite{Lutkenhaus99,BrassardSalvail93}. In a
bidirectional algorithm information can flow both ways, and Alice
can use the feedback from Bob to determine what additional
information she should provide him.  This makes it easier to
construct efficient algorithms.  These two classes can be further
subdivided into two subclasses, one for algorithms which discard
errors and one for those which correct them.  Discarding errors is
usually done in oder to prevent additional side information from
leaking to Eve. Algorithms which correct errors allow for this
additional flow of side information and account for it during
privacy amplification. Since privacy amplification is typically a
very efficient process, algorithms which correct the errors tend
to perform better

The communication rate in QKD strongly depends on the type of
error correction algorithm used.  In order to get an estimate of
this rate we must at least decide on which of the four subclasses
the algorithm belongs to. Since we are interested in practical
systems, and because efficiency is very important in quantum key
distribution, we will assume that the algorithm is bidirectional
and corrects the errors. An example of such an algorithm can be
found in~\cite{BrassardSalvail93}.

\subsection{Privacy amplification} \label{PrivAmp}

After error correction, Alice and Bob share an error free string
$X$.  Eve has also potentially obtained at least partial
information about this string from attacks on the raw quantum
transmission and side information leaked during error correction.
In~\cite{Fuchs97} it is shown that even with a measured error rate
of $1\%-5\%$ a non-negligible amount of information on string $X$
could have been revealed. Thus $X$ cannot by itself be used as a
key. However, through the method of generalized privacy
amplification~\cite{BennettBrassard95}, the string $X$ can be
compressed to a shorter string $K$ over which any eavesdropper has
only a negligible amount of information. The amount of compression
needed depends on how much information may have been compromised
during the previous phases of the transmission.

To do privacy amplification Alice picks a function $g$ out of a
universal class of functions $\mathcal{G}$ which map all $n$ bit
strings to $r$ bit strings where $r<n$
(see~\cite{BennettBrassard95} for more details).  Once $g$ has
been picked and publicly announced both parties calculate the
string $K=g(X)$, which serves as the final key. This key is
considered secure if Eve's mutual information on $K$, defined
as~\cite{Cover}
  \begin{equation}
    I_E(K;GV) = H(K) - H(K|GV),
  \end{equation}
is negligibly small, where $G$ is the random variable
corresponding to the choice of function $g$ and $V$ is all the
information available to Eve.

An important quantity in the analysis of privacy amplification is
the collision probability defined as
  \begin{equation}
    P_c(X) = \sum_x p^2(x).
  \end{equation}
One can show that the conditional entropy $H(K|G)$ is bounded
by~\cite[Thm. 3] {BennettBrassard95}
  \begin{equation} \label{Hbound}
    H(K|G) \ge r - \frac{2^r}{\ln 2}P_c(X)
  \end{equation}
This theorem can be applied to conditional distributions as well,
which leads to
  \begin{equation}
    H(K|G,Z=z) \ge r - \frac{2^r}{\ln 2}P_c(X|Z=z),
  \end{equation}
where $P_c(X|Z=z)$ is just the collision probability of the
distribution $p(x|Z=z)$.  By averaging both sides of the above
equation we get
  \begin{equation} \label{EntBound}
    H(K|GZ) \ge r - \frac{2^r}{\ln 2}\avg{P_c(X|Z=z)}_z,
  \end{equation}
where
  \begin{equation}
    \avg{P_c(X|Z=z)}_z = \sum_{z} p(z) P_c(X|Z=z)
  \end{equation}
is the average collision probability.  This is a quantity of
central importance in privacy amplification. In the case of
individual attacks, the i'th bit in $Z$ depends only on the i'th
bit in $X$.  Under these circumstances the average collision
probability factors into the product of the average collision
probability of each bit.  Thus,
  \begin{equation} \label{CPfactor}
    \avg{P_c(X|Z=z)}_z = (p_c)^{n},
  \end{equation}
where $n$ is the number of bits in string $X$ and
  \begin{equation} \label{SingleBitPc}
    p_c = \sum_{\alpha=0,1} \sum_{\beta=1}^k
    \frac{p^2(\alpha,\beta)}{p(\beta)}.
  \end{equation}
In the above expression $\alpha$ sums over the possible values of
a single bit in Alices string and $\beta$ sums over the possible
measurement outcomes of the probe, which are enumerated from $1$
to $k$. Suppose that we are able to come up with a bound of the
form $-\log_2\avg{P_c(X|Z=z)}_z~\ge~c$. If we set $r=c-s$, where
$s$ is a security parameter chosen by Alice and Bob, then
(\ref{EntBound}) leads to
  \begin{equation} \label{IeBound1}
    I_E(X;Z)~\le~2^{-s}/\ln2.
  \end{equation}
Thus, a bound on the average collision probability tells how short
we should make string $K$.

Before concluding this review of the main concepts in privacy
amplification we would like to make a few comments on the notion
of security in QKD. As stated above we consider the key secure if
the mutual information is very small. One might raise concerns
about this definition of security. The mutual information can be
interpreted as the average number of bits Eve will obtain on the
final key. In any given experiment it is possible that Eve can
obtain significantly more bits than the average, but this happens
with small probability. Perhaps a more satisfactory notion of
security would be a statement of the form, with probability no
greater than $\varepsilon$ Eve obtains no more than $\varsigma$
bits of information on the final key.  The mutual information is
an important quantity because it allows us to make such a
statement.  A simple method for doing this is to use the Markov
bound
  \begin{equation}
    P(I \ge \varsigma) \le \frac{I_E(K;GUZ)}{\varsigma},
  \end{equation}
where $I$ is the actual number of bits of information Eve has
obtained.  Setting $\varsigma=1$ gives us a bound on the
probability that Eve obtains more than one bit of information on
the final string.  This may serve as a more convincing statement
of security than statements about the average.
Plugging~(\ref{IeBound1}) into the above expression shows that the
probability that Eve obtains more than one bit on the final key is
exponentially small in the security parameter $s$.

\subsection{Handling side information from error correction}
\label{SideInfo}

If the only information available to Eve comes from string $Z$,
which is obtained from attacks on the quantum transmission, then
the discussion in the previous section is sufficient.  But in the
case of bi-directional error correction Eve will also learn an
additional string $U$ which gives her more information about
Alice's key.  This side information must also be included in the
calculation.  We can apply the bound in (\ref{Hbound}) to the
conditional distribution $p(x|U=u,Z=z)$, which leads to
  \begin{equation} \label{GUZbound}
    H(K|G,U=u,Z=z) \ge  r - \frac{2^r}{\ln 2} P_c(X|U=u,Z=z).
  \end{equation}
We can then try to average both sides of the above expression but
doing this introduces additional complications. The random
variable $U$ introduces correlations between different bits in
strings $X$ and $Z$. Because of this the average collision
probability no longer factors into the product of individual bits,
as in (\ref{CPfactor}).  This makes the problem of finding a bound
on the average collision probability significantly more difficult.
This problem has been previously investigated in~\cite{Cachin97},
where several bounds on the collision probability $P_c(X|Z=z,U=u)$
were derived. But the extension of this work to the average
collision probability involves a few subtleties, which we deal
with in Appendix~\ref{InfoBound}.  In this appendix we show that
if we set
  \begin{equation} \label{FinalString}
    r = n\tau - \kappa - t - s,
  \end{equation}
where
  \begin{equation} \label{TauDef}
    \tau=-\log_2p_c,
  \end{equation}
$\kappa$ is the number of bits in message $U$, $n$ is the length
of the error corrected key, and both $s$ and $t$ are security
parameters chosen by Alice and Bob, then
  \begin{equation} \label{SecurityStatement}
    I_E \le 2^{-t}r + \frac{2^{-s}}{\ln 2}.
  \end{equation}
This bound on Eve's information is still exponentially small in
the security parameters, and only involves the collision
probability averaged over her measurements on the quantum
transmission.

\subsection{Communication rates for BB84} \label{BB84rate}

The main effort in proving security of quantum key distribution
against individual attacks comes down to finding a bound for
$p_c$, defined in (\ref{SingleBitPc}).  This bound should come
from the laws of quantum mechanics.  The quantity $p_c$ has been
extensively studied by L\"{u}tkenhaus in the context of the BB84
protocol~\cite{Lutkenhaus99}.  In this work several bounds are
derived using the model of a POVM, which Eve performs on each
quantum transmission. L\"{u}tkenhaus points out that it is better
not to discard any signals, even ambiguous dual fire events, in
order not to open up a security loophole. Monitoring the dual fire
events can be very useful to prevent Eve from sending Bob
additional photons. This is precisely why we picked passive
modulation, where the $50/50$ beamsplitter randomly partitions all
incoming photons. If Eve uses more than one photon she will cause
dual fire events with very high probability. By keeping track of
these events and dealing with them during error correction, Alice
and Bob can make it unfavorable for Eve to employ such tactics.
One convenient way to keep track of dual fire events is to define
a disturbance measure
  \begin{equation} \label{disturbance}
    \epsilon = \frac{n_{err} + w_Dn_D}{n_{rec}},
  \end{equation}
where $n_{rec}$, $n_{err}$, and $n_D$ are the number of error
corrected bits, error bits, and ambiguous dual fire events
respectively, and $w_D$ is an independently chosen weighting
parameter.  The value of $w_D$ should be made sufficiently large
so that it is to Eve's advantage to only use single photons. It is
shown in~\cite{Lutkenhaus99} that if the passive modulation scheme
is used and $w_D$ is set to $1/2$ the collision probability can be
bounded by
  \begin{equation} \label{CollisionProb}
    p_c \le \frac{1}{2} + 2 \epsilon - 2\epsilon^2.
  \end{equation}
Throughout our calculations we will assume that dual fire events
are very rare.  In this limit we have $\epsilon=e$, where $e$ is
the error rate.  This approximation is usually reasonable, and is
extremely good in the limit of large loss which is what we are
mainly interested in.

The above bound on the collision probability is valid for the BB84
protocol only if the photon source never injects more than one
photon into the channel.  However, practical photon sources
sometimes inject a multi-photon state. These states are vulnerable
to photon splitting attacks where Eve splits one of the photons
from the pulse and leaves the remaining ones undisturbed.  This
can be done with a Jaynes-Cummings type
interaction~\cite{Lutkenhaus00}. She can store this photon
coherently until the measurement basis is revealed, after which
she learns the exact value of the bit. The most powerful types of
photon splitting attacks will also presume that Eve has access to
lossless optical fibers.  This allows her to transmit the
remaining photons from a multi-photon states with unity
probability.  At the same time she blocks off a fraction of the
single photon states while conserving the overall transmission
rate, giving her complete knowledge over a larger fraction of the
sifted key.  The extension of security for realistic sources which
sometimes create a multi-photon state is given
in~\cite{Lutkenhaus00}. Each multi-photon state can in principle
result in a collision probability of one while producing no
errors.  This can be accounted for by first defining the parameter
  \begin{equation} \label{BetaDef}
    \beta = \frac{n_{rec} - n_m}{n_{rec}}
  \end{equation}
where $n_m$ is the number of transmissions in which more than one
photon was injected into the channel. The definition of $\tau$ in
(\ref{TauDef}) should then be modified to~\cite{Lutkenhaus00}
  \begin{equation} \label{TauMulti}
    \tau = - \beta\log_2 p_c(e/\beta),
  \end{equation}
where $p_c$ is the collision probability of a single photon state
which is bounded by (\ref{CollisionProb}).

We can now put all the previously discussed elements together to
calculate the communication rate of the BB84 protocol with both
ideal and realistic photon sources.  An ideal source generates
exactly one photon every clock cycle of the experiment.  Such
sources are beyond current technological capabilities, so most
implementations of BB84 use photon sources based on attenuated
coherent light.  In such schemes the number of photons $N$ in
each pulse follows a poisson distribution
  \begin{equation}
    P(N=j) = e^{-\nbar}\frac{\nbar^j}{j!}
  \end{equation}
where $\nbar$ is the average number of photons per pulse.  The
average photon number should be made sufficiently low so that the
the probability of injecting more than one photon into the channel
is small.  At the same time $\nbar$ cannot be made small without
having a larger fraction of the pulses contain zero photons. The
average number should be chosen carefully to balance both effects.

After the source emits a pulse its polarization is set by an
electro-optic modulator, and the signal is injected into the
channel.  It is detected by Bob with some probability which
depends on fiber losses, the quantum efficiency of Bob's detector,
and any other loss mechanism in the system.  We assume that the
channel transmission is an exponentially decaying function of
distance. Thus, the channel transmission $T_F$ can be written as
  \begin{equation} \label{Tparam}
    T_F = 10^{-(\sigma L/10)},
  \end{equation}
where $\sigma$ is the loss coefficient.  As previously discussed,
we combine all losses from the channel, detectors, and optics into
one beamsplitter with transmission
  \begin{equation} \label{alphaL}
    \Trans = \eta T_F
  \end{equation}
and one loss mode $c$, which is shown in Figure~\ref{BB84}. The
subscript $L$ is used to denote that the loss is a function of
distance. The factor $\eta$ accounts for all constant losses in
the system such as detector inefficiency and reflection loss from
optics.

In a practical system detection events can also arise from dark
counts in the detection unit. The probability that a detection
event occurs can be written as
  \begin{equation}
    p_{click} = p_{signal} + p_{dark} -
    p_{signal} p_{dark},
  \end{equation}
where $p_{signal}$ is the probability that the detector registers
a count due to a photon, and $p_{dark}$ is the probability it
registers a dark count. If the probability of a simultaneous
signal and dark count event is negligibly small we can write
  \begin{equation}
    p_{click} \approx p_{signal} + p_{dark}.
  \end{equation}
The expression $p_{signal}$ can be written as
  \begin{equation} \label{PsignalDef}
    p_{signal} = \sum_{i=1}^{\infty} p(i) (1 -
    (1-\Trans)^i),
  \end{equation}
where $p(i)$ is the probability the source generated $i$ photons.
For an ideal source $p(1)=1$ and we have the rather trivial result
that $p_{signal}=\Trans$.  For a poissonian light source we can
derive the closed form solution
  \begin{equation}
    p_{signal} = 1 - e^{-\Trans \nbar}.
  \end{equation}

We assume that all detectors have the same dark count rate, and
define $d$ as the probability of having a dark count within the
measurement time window.  If we neglect the events where more than
one dark count is detected in a clock cycle then
  \begin{equation}
    p_{dark} = 4d.
  \end{equation}
Thus, the error rate $e$ is given by
  \begin{equation}
    e = \frac{p_{dark}/2+\mu p_{signal}}{p_{click}}.
  \end{equation}
where $\mu$ takes into account the error rate of the signal
photons, which may result from imperfect polarizing optics or
channel distortion.  The expected number of bits in the error
corrected key, $n_{rec}$, is given by
  \begin{equation} \label{NsifDef}
    n_{rec} = \frac{n_{tot}p_{click}}{2},
  \end{equation}
where $n_{tot}$ is the total number of pulses sent by Alice.

We now need estimate of $\beta$ defined in (\ref{BetaDef}). In the
limit of long strings $\beta$ becomes
  \begin{equation}
    \beta = \frac{p_{click} - p_m}{p_{click}}.
  \end{equation}
where $p_m$ is the probability that more than one photon is
injected into the channel by a pulse.  An ideal source never emits
more than one photon we so $p_m=0$. For poisson light we have
  \begin{equation}
    p_m = 1 - (1+\nbar)e^{-\nbar}.
  \end{equation}

Finally, we need to take into account the side information leaked
by the correction.  We define the function $f(e)$ in the same way
as was done in~\cite{Lutkenhaus00}.  This function determines how
far off from the Shannon limit the error correction algorithm is
performing. Thus,
  \begin{equation}
    \lim_{n_{rec} \to \infty}\frac{\kappa}{n_{rec}} = -f(e)\left[e
    \log_2 \left( e \right) + \left( 1-e \right)
    \log_2 \left( 1 - e \right)\right].
  \end{equation}
The value of $f(e)$ can be determined by benchmark tests.  Such
tests have been performed for the algorithm given
in~\cite{BrassardSalvail93}, and values for $f(e)$ taken from this
experiment are shown in table~\ref{benchmark}.  We interpolate
these values to determine $f(e)$ for intermediate values of $e$.

  \begin{table}
     \centering
     \begin{tabular}{@{\hspace{.5in}}ll@{\hspace{.5in}}}
      e & f(e) \\  \hline
      0.01 & 1.16 \\
      0.05 & 1.16 \\
      0.1 & 1.22 \\
      0.15 & 1.35
    \end{tabular}
    \caption{Benchmark performance of error correction algorithm.}
    \label{benchmark}
  \end{table}

We can now put everything together. Using (\ref{FinalString}) one
finds
  \begin{equation}
    r = n_{rec} \left( \beta\tau(e/\beta) -
    \frac{\kappa}{n_{rec}}  \right) - t - s,
  \end{equation}
with the expression for $n_{rec}$ given in (\ref{NsifDef}). One
can then define the communication rate $R_{BB84}$ as
  \begin{eqnarray*}
    R_{BB84} & = & \lim_{n_{tot} \to \infty} \frac{r}{n_{tot}} \\
    & = & \frac{p_{click}}{2} \left\{ \beta\tau(e/\beta) + f(e)
    \left[ e \log_2 \left( e \right) + \left( 1-e \right) \log_2 \left( 1 - e
    \right) \right] \right\}.
  \end{eqnarray*}
This is the normalized rate per clock pulse, which can be
multiplied by the clock rate of the source to get the actual bit
rate.

\section{Security of the Ekert protocol}
\label{rates}

In this section we tackle the problem of proving security for the
Ekert protocol.  As shown in the previous section, this involves
finding an upper bound on $p_c$ given in (\ref{SingleBitPc}) using
the laws of quantum mechanics. We derive this bound  by allowing
Eve to generate any density matrix $\rho_{abe}$ which she will
share with Alice and Bob.  We then calculate the communication
rate for the Ekert protocol in the presence of detector dark
counts and channel losses.  This is done for both an ideal source
which creates exactly one entangled pair per clock cycle, as well
as a more practical source based on parametric down-conversion.

\subsection{Proof of security against individual attacks}

We begin the proof of security by defining the Hilbert space over
which Alice and Bob make their measurements. Alice and Bob's
signal are assumed to be distinguishable by their spatial and
momentum states.  Let us assume that Alice's photon is in mode $a$
and Bob's photon is in mode $b$.  The most general photon signal
in these two modes can be written as a linear superposition of the
eigenstates
  \begin{equation}
    \|\psi>_{ab} =
    \|n_1,n_2>_a\|n_3,n_4>_b
  \end{equation}
where $n_1$, and $n_2$ are the number of photons in the $x$ and
$y$ polarization of mode $a$, and $n_3$ and $n_4$ are the number
of photons in the $x$ and $y$ polarization of mode $b$.

Eve is allowed to pick any density matrix $\rho_{abe}$ which
represents some entangled state of her probe and the signals
transmitted to Alice and Bob.  She can send any number of photons
she wishes, or a coherent superposition of photon numbers.  We
first define the disturbance in the same way as
(\ref{disturbance}). The only difference is that now both Alice
and Bob could see a dual-fire event, so $p_D$ is interpreted as
the probability that either sees such an event.

We assume that Eve has complete control over whether a photon is
detected or not, so she can provide any density matrix
$\rho_{abe}$ of her choosing.  Our first step is to show that
Eve's best strategy is to keep track of how many photons she is
sending to both receivers.  This is done in
Appendix~\ref{coherence}, where we first show that off diagonal
terms in $\rho_{abe}$ which couple different photon number states
sent to Alice and Bob do not contribute to any measurement
results. This is simply a consequence of the detection apparatus
used by Alice and Bob, which is composed of passive linear optics
and photon counters. These elements alone cannot distinguish
between a coherent superposition and random mixture of photon
number states. By keeping track of how many photons she is sending
Eve destroys these off diagonal terms.  But since they never
factor into the measurements she can do this without effecting the
disturbance, and knowing the number of photons received can
potentially refine her attack.

The main consequence of the above result is that the most general
density matrix $\rho_{abe}$ can be assumed to be in the block
diagonal form
  \begin{equation}
    \rho_{abe} =  \sum_{i,j=1}^{\infty} \rho_{abe}^{(ij)}
  \end{equation}
where $\rho_{abe}^{(ij)}$ spans the subspace where Alice is sent
$i$ photons and Bob is sent $j$ photons.  Furthermore, because Eve
knows how many photons were sent her collision probability can be
broken up into different photon number contributions as
  \begin{equation}
    p_c = \sum_{i,j = 1}^{\infty}
    \frac{p_{rec}^{(ij)}}{p_{rec}}p_c^{(ij)},
  \end{equation}
where
  \begin{equation}
    p_c^{(ij)} = \sum_{m\in M^{(ij)},\psi} \frac{1}{p_{rec}^{(ij)}}
    \frac{p^2(\psi,m)}{p(m)}.
  \end{equation}
The set $M^{(ij)}$ is defined as the set of all measurement
results on Eve's probe if she sent $i$ photons to Alice and $j$ to
Bob, and $p_{rec}^{(ij)}$ is the probability that the signal
component $\rho_{abe}^{(ij)}$ enters the error corrected key. We
can similarly break up the disturbance measure $\epsilon$ into
different photon number contributions as
  \begin{equation} \label{disturbance2}
    \epsilon = \sum_{ij} \frac{p_{rec}^{(ij)}}{p_{rec}}
    \frac{p_{err}^{(ij)}+ w_Dp_D^{(ij)}}{p_{rec}^{(ij)}} =
    \sum_{ij} \frac{p_{rec}^{(ij)}}{p_{rec}} \epsilon^{(ij)}.
  \end{equation}
In the above expression $p_{err}^{(ij)}$ is the probability that
$\rho_{abe}^{(ij)}$ enters the sifted key as an error, and
$p_D^{(ij)}$ is the probability that it causes a dual fire event.

Our next step is to investigate the term $p_c^{(11)}$. In
Appendix~\ref{OnePhoton} we show that this term is bounded by
  \begin{equation}
    p_c^{(11)} \le \frac{1}{2} + 2 \epsilon^{(11)} -
    2\left(\epsilon^{(11)}\right)^2.
  \end{equation}
Once this is established, we show in Appendix~\ref{MultiPhoton}
that if the weighting parameter $w_D$ in (\ref{disturbance2}) is
set to $1/2$ than Eve's optimal strategy is to only send one
photon to Alice and Bob.  Given that this is the optimal strategy
one is led directly to the result
  \begin{equation}
    p_c \le \frac{1}{2} + 2 \epsilon -
    2\epsilon^2.
  \end{equation}
which is exactly the same collision probability given in
(\ref{CollisionProb}).

Our proof makes no idealized assumptions about the entangled
photon source, such as that it only emits one pair of photons per
clock cycle.  As a matter of fact the source never enters the
picture since we assume that Eve blocks it out and replaces it
with the best possible source allowed by the laws of quantum
mechanics. The error rate and dual fire rate alone are enough to
put a bound on her collision probability.

\subsection{Ideal entangled photon source}

Before analyzing practical entangled photon sources for the Ekert
protocol we will first analyze the simpler case of an ideal
entangled photon source. This source is assumed to create exactly
one pair of photons per clock cycle, whose quantum state is given
by
  \begin{equation} \label{MES}
    \|\psi_+> = \frac{1}{\sqrt{2}} \left( \|HV> + \|VH> \right).
  \end{equation}
Although proposals for creating such a source
exist~\cite{BensonSantori00}, we do not know of any successful
implementations of this proposal to date. Nevertheless, this
simplified analysis will set the groundwork for the analysis of
entangled photon sources based on parametric down-conversion.

When doing two photon experiments, one is no longer interested in
individual detection events.  Instead, one looks for coincidence
events where two detectors simultaneously fire.   We separate the
coincidence probability into two parts, $p_{true}$ is the
probability of a true coincidence from a pair of entangled
photons, and $p_{false}$ is the probability of a false coincidence
which for an ideal source can only occur from a photon and dark
count, or two dark counts.  We write
  \begin{equation} \label{pcoin}
    p_{coin} = p_{true} + p_{false},
  \end{equation}
where once again the probability of a simultaneous true and false
coincidence are considered negligibly small.

First we need to decide where to put the source.  Using the
definition in (\ref{alphaL}), we set the source a distance $x$
from Alice and $L-x$ from Bob.  Then
  \begin{eqnarray*}
    p_{true} & = & \alpha_x \alpha_{L-x}, \\
    & = & \Trans,
  \end{eqnarray*}
and
  \begin{equation}
    p_{false}  =  \alpha_x \left( 1 - \alpha_{L-x} \right)
    4d(1-d)^3 + \alpha_{L-x} \left( 1 - \alpha_x \right) 4d(1-d)^2 +
    16d^2(1-d)^2.
  \end{equation}
It can be seen that the probability of a true coincidence does not
change with $x$, but the false coincidence rate does.  A simple
optimization shows that the false coincidence rate achieves a
minimum halfway between Alice and Bob.  At this optimal location
the false coincidence rate is
  \begin{equation}
    p_{false}  =  8\T2d + 16d^2,
  \end{equation}
where we keep only terms that are quadratic in $\T2$ and $d$.
Higher order terms can be ignored because they correspond to more
than one detection event at either receiver. The error rate $e$ is
  \begin{equation} \label{erate}
    e = \frac{p_{false}/2 + \mu p_{true}}{p_{coin}},
  \end{equation}
which then leads to an expression for the communication rate
$R_{Ekert}$
  \begin{equation} \label{EkertRate}
    R_{Ekert} = \frac{p_{coin}}{2} \left\{\tau(e) + f(e) \left[ e
    \log_2e + (1-e) \log_2 \left( 1 - e \right) \right] \right\}.
  \end{equation}

\subsection{Entangled photons from parametric down-conversion}

A more practical way of generating entangled photons is to use the
spontaneous emission of a non-degenerate parametric amplifier.
This technique, known as parametric down-conversion, is
extensively used to generate entanglement in polarization as well
as other degrees of freedom such as energy and momentum.
Parametric amplifiers exploit the second order non-linearities of
non-centrosymmetric materials.  These non-linearities couple three
different modes of an electromagnetic field via the interaction
Hamiltonian~\cite{WallsMilburn}
  $$
    H_I = i\hbar\chi^{(2)} Ve^{i(\omega-\omega_a - \omega_b) t}
    \hat{a}^{\dagger}\hat{b}^{\dagger} + \mbox{h.c.}
  $$
where modes $a$ and $b$ are treated quantum mechanically while the
third mode $Ve^{i\omega t}$ is considered sufficiently strong to
be treated classically.  The state of the field after the
nonlinear interaction is given by
  $$
    \|\psi> = \exp\left[ \frac{1}{i\hbar} \int_0^T H_I(t)
    dt \right]\|0>.
  $$
We assume the energy conservation condition,
$\omega=\omega_a+\omega_b$, which leads directly to
  $$
    \|\psi> = e^{\chi(\adag\bdag - \hat{a}\hat{b})}\|0>,
  $$
where the parameter $\chi$ depends on several factors including
the non-linear coefficient $\chi^{(2)}$, the pump energy, and the
interaction time. Using the operator identity~\cite{WallsMilburn}
 \begin{equation}\label{Disentangle}
    e^{\chi(\adag\bdag - \hat{a}\hat{b})} = e^{\Gamma\adag\bdag}
    e^{-g(\adag\hat{a} + \bdag\hat{b} + 1)}
    e^{-\Gamma\hat{a}\hat{b}},
  \end{equation}
where
  \begin{eqnarray*}
    \Gamma & = & \tanh\chi \\
    g & = & \ln\cosh\chi,
  \end{eqnarray*}
directly leads to the relation
  \begin{equation}
    \|\psi> = \frac{1}{\cosh \chi} \sum_{n=0}^\infty \tanh^n
    \chi \|n>_a \|n>_b
  \end{equation}
The above equation makes it clear that whenever a photon is
detected in one mode, the conjugate mode must also contain a
photon. In order to generate entanglement in polarization one
needs to create a correlation between the polarization of these
two modes.  This is typically done using non-colinear Type II
phase matching~\cite{KwiatMattle95}, which leads to the slightly
more complicated interaction
  $$
    H_I = i\hbar \chi^{(2)} A e^{i\omega t} \left( \hat{a}_{x}^{\dagger}
    \hat{b}_{y}^{\dagger} + \hat{a}_{y}^{\dagger}
    \hat{b}_{x}^{\dagger} \right) + \mbox{h.c.}
  $$
where $x$ and $y$ refer to the polarization of the photon. Since
all creation operators in the Hamiltonian commute, we can apply
(\ref{Disentangle}) to both mode pairs which directly leads to
  \begin{equation}
    \|\psi>
    =\frac{e^{\tanh\chi \left(\hat{a}_{x}^{\dagger}\hat{b}_{y}^{\dagger} +
    \hat{a}_{y}^{\dagger}\hat{b}_{x}^{\dagger}\right) }}{\cosh^2\chi}\|0>
  \end{equation}
In the limit of small $\chi$ one can make the approximation
  \begin{equation}
    \|\psi> \approx \sqrt{ 1 - 2\chi^2} \|0> + \sqrt{2}\chi \left( \|1>_{a_x}
    \|1>_{b_y} \|0>_{a_y} \|0>_{b_x} + \|0>_{a_x}
    \|0>_{b_y} \|1>_{a_y} \|1>_{b_x} \right)
  \end{equation}
Thus, a parametric down-converter creates an approximate Bell
state if $\chi$ is sufficiently small to ignore higher order
contributions.  But $\chi$ cannot be made small without
sacrificing the rate of down-conversion.

We want to calculate the probability $p_{coin}$ and the error rate
$e$ as a function of the parameter $\chi$, as well as the optical
losses and dark counts of the detectors. We begin by defining the
field operator
  \begin{equation} \label{WaveOp}
    \hat{\psi} = \frac{e^{\tanh\chi \left(\hat{a}_{x}^{\dagger}\hat{b}_{y}^{\dagger} +
    \hat{a}_{y}^{\dagger}\hat{b}_{x}^{\dagger}\right)
    }}{\cosh^2\chi}.
  \end{equation}
The beamsplitter model that we have introduced previously to
account for the losses becomes very useful here.  The
beamsplitters perform a unitary operation on the modes which is
given by
  \begin{eqnarray*}
    \hat{a}_\sigma & \to & \sqrt{\T2}\hat{a}_\sigma +
    \sqrt{1-\T2}\hat{c}_\sigma, \\
    \hat{b}_\sigma & \to & \sqrt{\T2}\hat{b}_\sigma +
    \sqrt{1-\T2}\hat{d}_\sigma, \\
  \end{eqnarray*}
where $\sigma$ represents polarization and the modes $c$ and $d$
are the reflected modes of the beamsplitter.  To determine the
state of the photons after the loss we first apply this
beamsplitter transformation.  To simplify the notation we define
another field operator
  $$
    \psi_{\rho\phi} = \hat{\rho}_x^{\dagger}\hat{\phi}_y^{\dagger}
    + \hat{\rho}_y^{\dagger}\hat{\phi}_x^{\dagger}
  $$
where $\rho$ and $\phi$ are any two independent modes.  Using this
definition, (\ref{WaveOp}) is transformed by the two beamsplitters
into
  $$
    \hat{\psi} = \frac{1}{\cosh^2\chi}\exp\left[{\tanh\chi \left(
    \T2\psi_{ab} + \sqrt{\T2\left( 1 - \T2
    \right)} \left( \psi_{ad} + \psi_{bc} \right) + \left( 1 -
    \T2 \right) \psi_{cd} \right)}\right].
  $$
We can expand this expression in terms of $\adag$ and $\bdag$ as
  \begin{eqnarray*}
    \hat{\psi} & = & \frac{1}{\cosh^2\chi}
    \exp\left[\tanh\chi \left( 1 - \T2 \right) \psi_{cd}\right]
    \left\{1 + \sum_{n=1}^{\infty} \frac{\tanh\chi^n}{n!}\left[
    n\sqrt{\T2(1-\T2)}\left(\psi_{ad} +
      \psi_{cb}\right)\right. \right.  \\
    & & \left. \left.
    \T2 \left( n\psi_{ab} + n(n-1)\left(1-\T2\right)\psi_{ad}
      \psi_{cb} \right)\right]
     + \psi_D \right\}
  \end{eqnarray*}
where $\psi_D$ is the wave operator which contains all the terms
that create more than one photon in either modes. It is now
necessary to operate on the vacuum and trace out over modes $c$
and $d$ to get the final density matrix. As shown in
Appendix~\ref{coherence} we can ignore any off diagonal terms that
couple different photon number states because they do not
contribute to the signal. We define the density matrix
$\rho_{\psi_+}$ as the two photon density matrix in which the
photons are in the entangled state $\|\psi_+>$ given in
(\ref{MES}). The matrices $\rho_0^a$ and $\rho_0^b$ represent a
zero photon vacuum state in mode $a$ and $b$ respectively. Finally
we define the matrices $\rho_u^a$ and $\rho_u^b$ as
  \begin{equation}
    \rho_u^{a,b} = \frac{I}{2},
  \end{equation}
where $I$ is the identity matrix.  The above matrices correspond
to an unpolarized photon in mode $a$ or $b$ respectively. After
tracing out loss modes $c$ and $d$ the density matrix becomes
  \begin{equation} \label{RhoSingle}
    \rho_{AB} = A \rho_{\psi_+} + B\rho_0^a\otimes\rho_0^b +
    C\left( \rho_u^a\otimes\rho_0^b +
    \rho_0^a\otimes\rho_u^b \right)  + D\rho_u^a\otimes\rho_u^b +
    \left( 1 - A - 2B - C - D \right)\rho_D,
  \end{equation}
where $\rho_D$ is the matrix which represents all the possible
states in which more than one photon is in either mode $a$ or $b$
after the losses.  The coefficients $A$,$B$,$C$, and $D$ are
  \begin{eqnarray}
    A & = & \frac{1}{\cosh^4\chi}\frac{2\alpha_{L/2}^2
    \tanh^2\chi}{\left( 1 - \tanh^2\chi \left( 1 -
    \alpha_{L/2}\right)^2 \right)^4}, \\ \label{real_coin}
    B & = & \frac{1}{\cosh^4\chi} \frac{1}{\left( 1 -
    \tanh^2\chi\left(1 - \alpha_{L/2}\right)^2 \right)}, \\
    C & = & \frac{1}{\cosh^4\chi} \frac{2\alpha_{L/2} \left(
    1- \alpha_{L/2} \right) \tanh^2\chi}{\left(1-\tanh^2\chi
    \left( 1 - \alpha_{L/2} \right)^2\right)^2}, \\
    D & = & \frac{1}{\cosh^4\chi}\frac{4\alpha^2\left( 1 -
    \alpha_{L/2} \right)^2 \tanh^4\chi}{1 - \tanh^2\chi
    \left( 1 - \alpha_{L/2} \right)^4}. \label{acc_coin}
  \end{eqnarray}
In the above expression, $A$ is the probability that Alice and Bob
share an entangled pair of photons.  This component on the signal
will be defined as a true coincidence, because it leads to error
free transmission.  The coefficient $B$ is then the probability
that neither receiver gets a photon, either because the source
failed to generate a pair or because all photons where lost.
Similarly, $C$ is the probability that one of the two receivers
gets a photon but the other does not.  In order for these signals
to be factored into the key the must be accompanied by dark
counts. Coefficient $D$ is the probability that both receivers get
a photon, but these photons are unpolarized and uncorrelated. Note
that $D$ is at least fourth order in $\tanh\chi$, indicating that
at least two pairs must be created in order for it to exist. The
intuitive explanation for the presence of this unpolarized
component is that when higher order number states are created, and
some of these photons are lost, the loss mode $c$ and $d$ play a
similar role to Eve. The photons in this mode can potentially
carry some information about the quantum state of the other
photons, and will thus result in decoherence. Since this component
of the signal causes a $50\%$ error, we can lump it into the
definition of a false coincidence.  Hence,
  \begin{eqnarray*}
    p_{true} & = & A \\
    p_{false} & = & 16d^2B + 8dC + D
  \end{eqnarray*}
The communication rate can be calculated by simply plugging  these
expressions into (\ref{erate}) and (\ref{EkertRate}).

\subsection{Calculations}

\begin{figure}
  \centering
  \epsfxsize = 4in
  \epsfbox{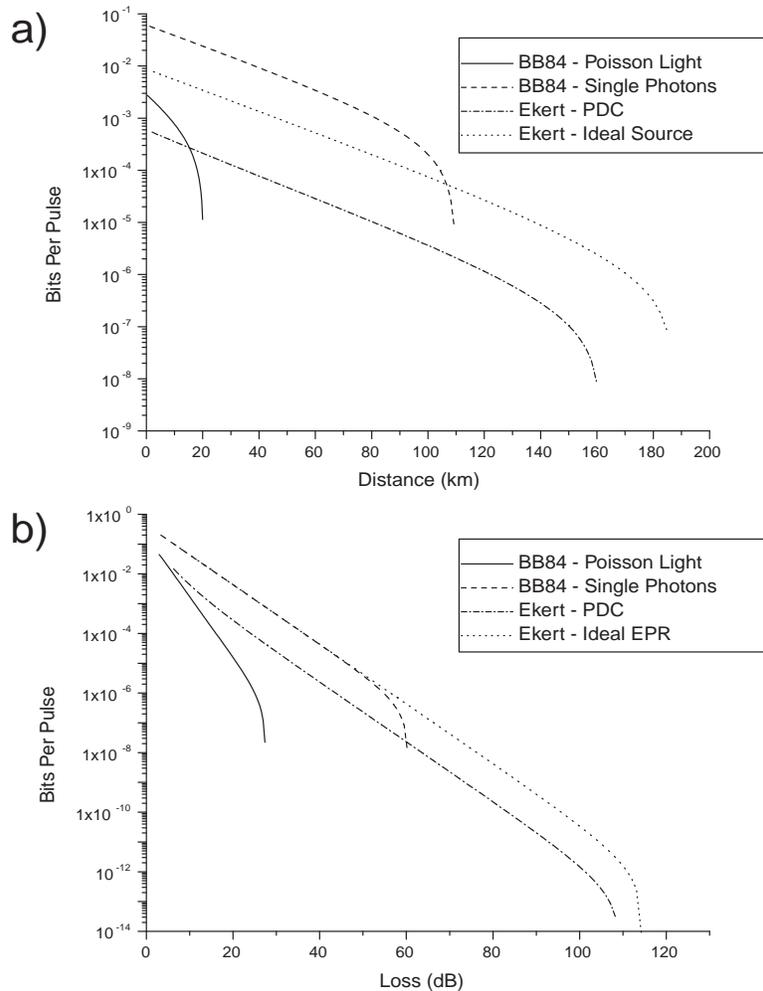}
  \caption{Comparison of communication rate for BB84 and Ekert
  protocol.  Plot (a) is for 1.5$\mu$m fiber optical communication
  experiment.  In this wavelength $\eta=0.18$, $d=5 \times
  10^{-5}$, and the channel loss $\sigma $ is set to 0.2dB/km. For
  the Ekert protocol the distance is the total separation between
  Alice and Bob.  Plot (b) shows calculated values for free-space
  quantum key distribution with visible photons.  The rate is
  plotted as a function of the total loss, including detector
  quantum efficiency.  The detectors are assumed to have a dark
  count rate of $d=5 \times 10^{-8}$.  For the Ekert protocol the
  loss is the total loss in both arms.}\label{simulation}
\end{figure}

We now use the previously derived equation to calculate the rate
of quantum key distribution using both the Ekert protocol and
BB84.  We perform simulations for fiber optical and free space key
distribution experiments.  For the fiber optical simulation we
look at the 1.5$\mu$m telecommunication window, while for free
space communication we focus on the visible wavelengths where
single photon counters tend to perform best.  In free space
communication the channel loss is no longer an exponential
function of distance. Instead, it is a complicated function which
results from atmospheric effects, beam diffraction, and beam
steering problems. Thus for free space we are more interested in
the rate as a function of the total loss rather than distance.

Figure~\ref{simulation} shows the calculation results, for both
BB84 and Ekert protocols with ideal and realistic sources. In plot
(a) of the figure we show results for fiber optical
communications. Using experimental values
from~\cite{BourennaneGibson99} we set the detector quantum
efficiency to $0.18$, $d=5\times 10^{-5}$, and the channel loss
$\sigma=0.2$dB/km. We also set the baseline error rate $\mu=0.01$,
and add an extra $1$dB of loss to account for losses in the
receiver unit. The curves corresponding to the Ekert protocol plot
the distance from Alice to Bob, with the source assumed halfway in
between. Plot (b) shows calculations for free space quantum key
distribution. The communication rate is plotted as a function of
the total loss, including the detector quantum efficiency.  In the
free space curves for the Ekert protocol we again put the source
halfway between Alice and Bob and plot the rate as a function of
the total loss in both arms.  The dark counts of the detectors are
set to $5\times 10^{-8}$.  In the curve for BB84 with a poisson
light source the average photon number $\nbar$ is a free
adjustable parameter. Similarly in the Ekert protocol with
parametric down-conversion we are free to adjust $\chi$.  For both
cases we numerically optimize the communication rate at each point
with respect to the adjustable parameter.

Each curve features a cutoff distance where the communication rate
quickly drops to zero. This cutoff is due to the dark counts,
which begin to make a non-negligible contribution to the signal at
some point.   However the two curves for the Ekert protocol
feature a much longer cutoff distance than the BB84 counterparts.
This is due partially to the absence of the photon splitting
attacks. But even when performing BB84 with ideal single photon
sources, which don't suffer from photon splitting attacks either,
the cutoff distance for the Ekert protocol is still significantly
longer. The reason for this is that in the Ekert protocol, a dark
count alone cannot produce an error. It must be accompanied by a
photon or another dark count, and thus is much less likely to
contribute to the signal. The difference in rates between the
ideal EPR source and the parametric down-converter can be
attributed to the interplay between coefficient $A$ in
(\ref{real_coin}), and coefficient $D$ in (\ref{acc_coin}).  Term
$A$ is the probability of a real coincidence, and increases with
$\chi$. Term $D$ on the other hand contributes to false
coincidences and increases with $\chi$ as well, but is of higher
order.  One cannot make $A$ arbitrarily large without getting an
increased contribution from $D$.  This leads to an optimum value
for $\chi$ which is less than one.

\section{Entanglement Swapping} \label{swap}

  \begin{figure}
    \centering
    \epsfxsize = 5in
    \epsfbox{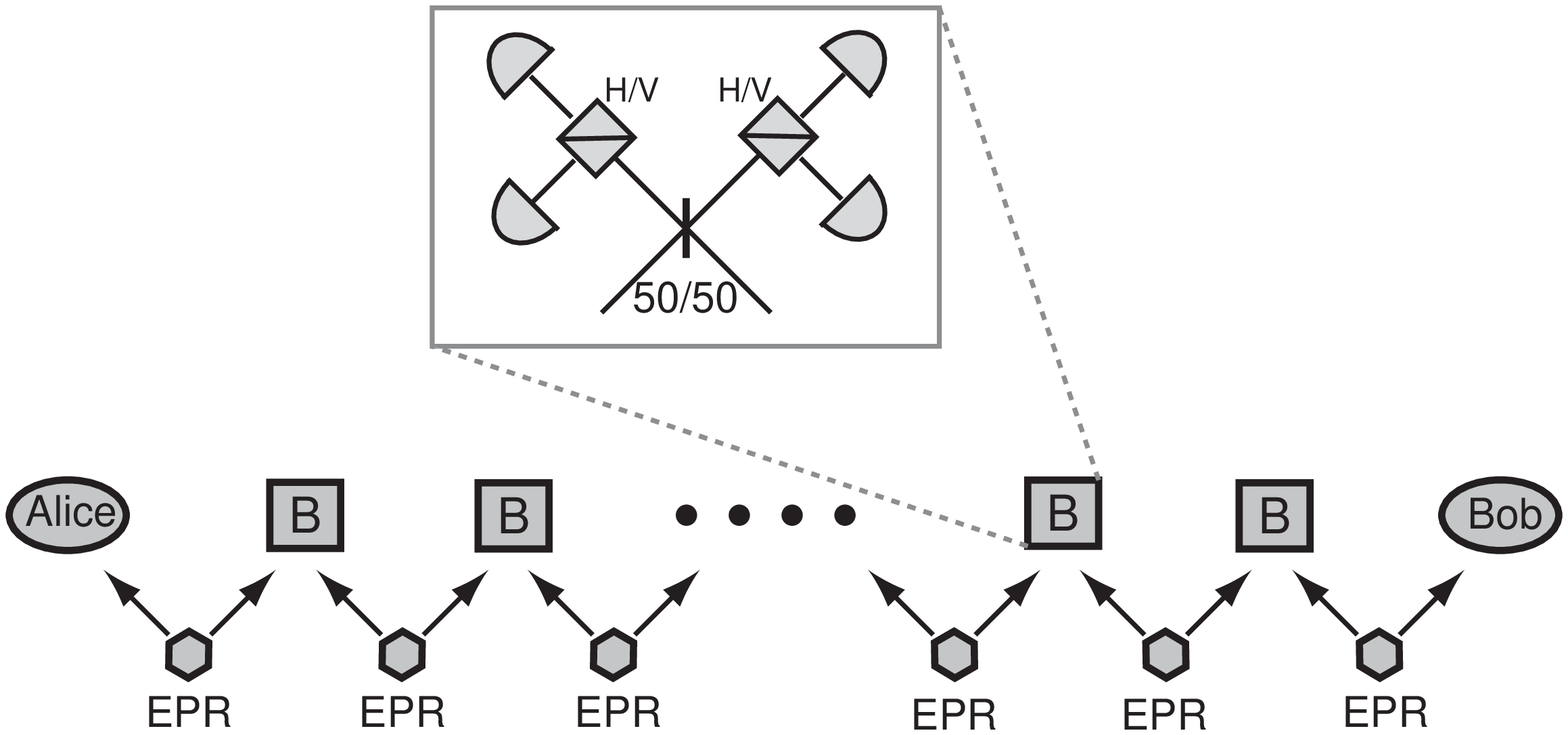}
    \caption{Experimental setup for quantum key distribution with
    entanglement swaps.}\label{repeater}
  \end{figure}

In this section we analyze a more complicated scheme based on
entanglement swapping. Figure~\ref{repeater} gives a diagram of
the proposed configuration.  A series of entangled photon sources,
which we assume to be ideal sources, are spread out an equal
distance distance apart from Alice to Bob. The sources are clocked
to simultaneously emit a single pair of entangled photons. Each of
the pair is sent to a corresponding Bell State Analyzer, whose
actions is to perform an entanglement swap.  If all the swaps have
been successfully performed, Alice and Bob will share a pair of
entangled photons. Experimental demonstrations of a single
entanglement swap can be found in~\cite{PanBouwmeester98}.
Entanglement swapping is a key element for quantum repeaters,
which use entanglement purification protocols to reliably exchange
quantum correlated photons between two
parties~\cite{BriegelDur98}. We show that even without such
protocols, using only linear optical elements, photon counters,
and a clocked source of entangled photons, swapping can enhance
the communication distance.

The key element to the scheme is the Bell Analyzer. Since we
restrict ourselves to passive linear elements and vacuum auxiliary
states we cannot achieve a complete Bell Measurement.  It has
recently been shown that Bell Analyzers based on only these
components cannot have better than a $50\%$
efficiency~\cite{CalsamigliaLutkenhaus00}.  One scheme which
achieves this maximum is shown on the inset of
Figure~\ref{repeater}.  This scheme will distinguish between the
states
  \begin{equation}
    \|\psi_\pm> = \frac{1}{\sqrt{2}} ( \|HV> \pm \|VH> )
  \end{equation}
but will register an inconclusive result if sent the states
  \begin{equation}
    \|\phi_\pm> = \frac{1}{\sqrt{2}} ( \|HH> \pm \|VV> )
  \end{equation}

The state generated by the entangled photon sources is assumed to
be $\|\psi_+>$.  Considering only a single swap, we can write
  \begin{equation}
    \|\psi_+>_{12} \|\psi_+>_{34} = \frac{1}{2}\left[
    \|\psi_+>_{23}\|\psi_+>_{14}-\|\psi_->_{23}\|\psi_->_{14} +
    \|\phi_+>_{23}\|\psi_+>_{14}-\|\phi_->_{23}\|\phi_->_{14}\right]
  \end{equation}
The above expression makes it clear that a Bell measurement on
photons $2$ and $3$ leaves photons $1$ and $4$ in an entangled
state, and the measurement result tells which one.  After $N$ such
Bell measurements photon $1$ and $2N$ will be entangled, and the
$N$ Bell measurement results will allow Alice and Bob to know
which entangled state they share.  Knowledge of this state allows
them to do entangled photon key distribution and interpret their
data correctly.  Since our Bell analyzer has an efficiency of only
$50\%$, in the best possible case we will pay a price of $2^{-N}$
in communication rate.

Consider the single swap. We will define $\alpha$ to be the
detection probability for each photon.  The probability that both
photon $2$ and $3$ reach the Bell analyzer and are successfully
projected is
  \begin{equation}
    p_{swap}^{true} = \frac{1}{2} \alpha^2
  \end{equation}
If a photon is lost in the fiber or due to detector inefficiency
the Bell analyzer may still indicate that a Bell measurement has
been performed due to detector dark counts.  The probability of
this happening is
  \begin{equation}
    p_{swap}^{false} = 6\alpha d + 12d^2.
  \end{equation}
Defining the factor
  \begin{equation}
    g = \frac{p_{swap}^{true}}{p_{swap}^{true}+p_{swap}^{false}}
  \end{equation}
it is straightforward to show that, given the Bell analyzer
registered a successful Bell measurement, the density matrix of
photons $1$ and $4$ is given by
  \begin{equation}
    \rho_{14} = g\rho_{\psi_\pm} + (1-g)\frac{I}{4}
  \end{equation}
where $\rho_{\psi_{\pm}}$ is the pure state $\|\psi_+>$ or
$\|\psi_->$ depending on the measurement result.

For the case of $N$ entanglement swaps the detection probability
for each photon is
  \begin{equation}
    \alpha =  \eta 10^{-\frac{\sigma L}{10 (2N+2)}},
  \end{equation}
where $L$ is the distance from Alice to Bob.  It is again
straightforward to show that after $N$ swaps, the state of photon
$1$ and $2N$ is
  \begin{equation}
    \rho_{1,2N} = g^N\rho_{\psi_\pm} + (1-g^N)\frac{I}{4}
  \end{equation}
and the probability that all $N$ bell measurements registered a
successful result is
  \begin{equation}
    p_{Bell} = (p_{swap}^{true}+p_{swap}^{false})^N
  \end{equation}
We then have
  \begin{eqnarray*}
    p_{true} & = & p_{Bell}^N g^N\alpha^2 \\
    p_{false} & = & p_{Bell}^N (8\alpha d + 16d^2 + (1-g^N)\alpha^2)
  \end{eqnarray*}
These can be plugged into (\ref{erate}) and (\ref{EkertRate}) to
get the final communication rate.

  \begin{figure}
    \centering
    \epsfxsize = 5in
    \epsfbox{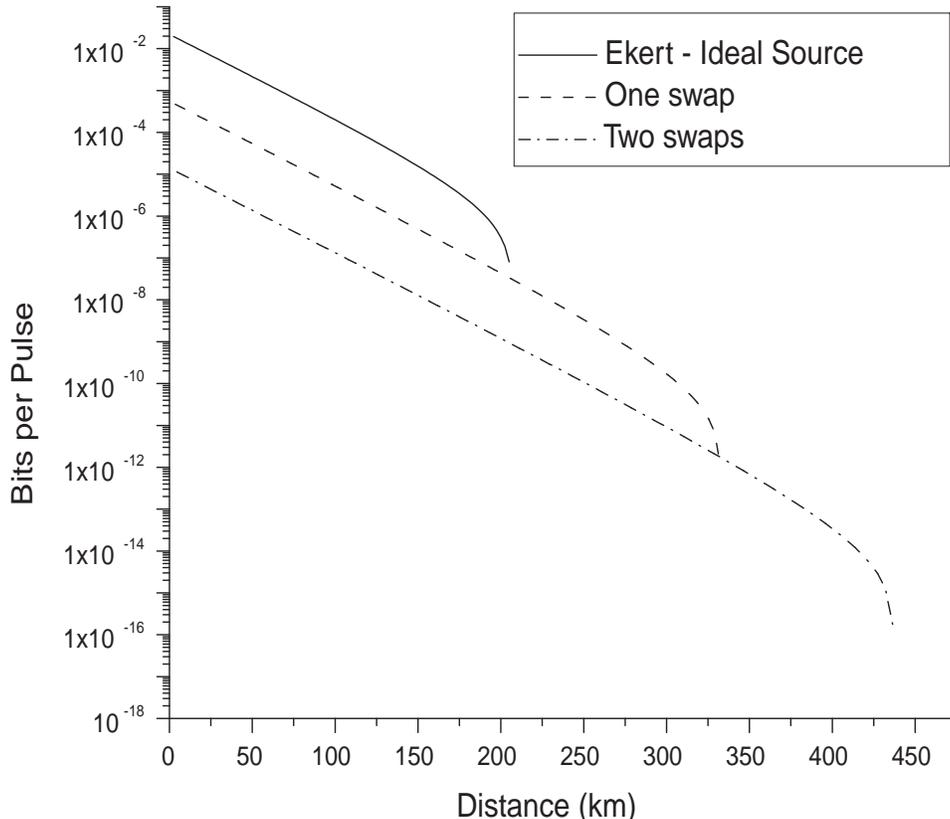}
    \caption{Comparison of Ekert protocol with regulated EPR source, one swap scheme,
    and two swap scheme.  Fibers and detectors are taken for the $1.5\mu$m window.} \label{SwapFig}
  \end{figure}
In Figure~\ref{SwapFig} we show a comparison between the Ekert
protocol with an ideal entangled photon source, a one swap scheme,
and a two swap scheme using a fiber optic channel at $1.5\mu$m.
The swaps result in a longer cutoff distance which can lead to
longer communication ranges.  It should be noted however that at
these distances the natural fiber loss is substantial and will
lead to very slow communication rates.  It is unclear whether
swapping will lead to a practical form of quantum key
distribution, but a single swap could be useful for very long
distance QKD.

\section{Discussion} \label{discussion}

A standard quantum key distribution experiment involves three
steps: raw quantum transmission, error correction, and privacy
amplification.  In the first two steps, Alice and Bob exchange a
key which is partially secure.  The purpose of the third step is
to compress this partially secure key to a shorter key which is as
secure as desired.  The natural question to ask is, what is the
relationship between the length of the final key and its security?
This question has already been answered in previous work for the
BB84 protocol with the restriction that Eve is only allowed to
attack each bit individually.  In this paper we provide a similar
answer for the Ekert protocol.

In summary, suppose that two parties wish to communicate using the
Ekert protocol.  They can then ensure that their final key is
secure against general individual attacks by performing the
following.  After error correction, they calculate the disturbance
measure which is defined as
  $$
    \epsilon = \frac{n_{err} + n_D/2}{n_{rec}},
  $$
where $n_{rec}$ is the number of bits in the error corrected
string, $n_{err}$ is the number of error bits, and $n_D$ is the
number of dual-fire events in which more than one photon counter
was triggered.  They next calculate the parameter $\tau$, which is
given by
  $$
    \tau = -\log_2 \left( \frac{1}{2} + 2 \epsilon - 2 \epsilon^2
    \right).
  $$
Using the techniques of generalized privacy amplification
discussed in~\cite{BennettBrassard95}, they compress their key to
a shorter key of length $r$ given by
  $$
    r = \tau n_{rec} - \kappa - s - t,
  $$
where $\kappa$ is the number of bits exchanged during error
correction and $t$ and $s$ are independent security parameters
chosen by the two parties.  The choice of these parameters depends
on how much security is desired on the final key, which is
quantified by an upper bound on Eve's mutual information.  This
bound is explicitly given by
  $$
    I_E(K;GUZ) \le 2^{-t}r + 2^{-s}/\ln 2,
  $$
which is exponentially small in $s$ and $t$.

Using the above results we compared the performance of BB84 and
Ekert for both ideal and practical sources.  We investigated
fiber-optic as well as free-space key distribution scenarios. The
Ekert protocol was shown to have significantly better performance
at longer distance provided that the source can be placed midway
between the two communicating parties.  This opens up the
possibility for communication lengths of up to $170$km, although
at low bit rates. The low bit rate is predominantly caused by the
fiber losses.

Finally, we analyzed a more complicated scheme based on
entanglement swaps using only linear optical components, photon
counters, and a clocked source of entangled photons. Entanglement
swapping can allow for even longer distance secure communication,
but at some point the natural loss of the fiber becomes so severe
that the communication rate is prohibitively slow.

\acknowledgments

The authors would like to thank Norbert L\"{u}tkenhaus for his
many comments and suggestions.

\begin{appendix}

\section{Information bounds on eavesdropping.}
\label{InfoBound}

In this appendix we show how to bound Eve's expected information
$I_E(K;GUZ)$ by the average collision probability
  \begin{equation}
    \avg{p_c(x|z)}_z = \sum_z p(z) P_c(X|Z=z),
  \end{equation}
where
  \begin{equation}
    P_c(X|Z=z) = \sum_x p^2(x|z).
  \end{equation}
Let $U$ and $Z$ be arbitrary, possibly correlated, random
variables over alphabets ${\mathcal U}$ and ${\mathcal Z}$
respectively. Let $|\cdot|$ denote the cardinality of a given set.
Let $t>0$ be a security parameter chosen by Alice and Bob and
define set $A$ as
  \begin{equation}
    A = \left\{ (u,z)\in \left( {\mathcal U}, {\mathcal Z}
    \right): p(u|z) \ge \frac{2^{-t}}{\left| {\mathcal U} \right|}
    \right\}.
  \end{equation}
Defining $A^c$ as the complement of set $A$ we have
  \begin{eqnarray*}
    P(A^c) & = & \sum_{(u,z)\in A^c} p(u,z) \\
    & = & \sum_{(u,z)\in A^c} p(u|z)p(z) \\
    & \le & \frac{2^{-t}}{|{\mathcal U}|} \sum_{u\in{\mathcal U},z\in{\mathcal Z}} p(z) \\
    & = & 2^{-t}.
  \end{eqnarray*}
Thus with probability of at least $1-2^{-t}$ the combined string
$(U,Z)$ take a value in $A$.  Then for another random variable $X$
  \begin{eqnarray*}
    \avg{P_c(X|Z=z)}_z & = & \sum_{z\in{\mathcal Z}} p(z) \sum_{x} p^2(x|z) \\
    & = & \sum_{z\in{\mathcal Z}} p(z) \sum_{x} \left(\sum_{u\in{\mathcal U}} p(u|z)p(x|uz)
    \right)^2
    \\
    & \ge & \sum_{z\in{\mathcal Z}} p(z) \sum_x \sum_{u\in{\mathcal U}} p^2(u|z) p^2(x|uz) \\
    & = & \sum_{z \in {\mathcal Z} , u \in {\mathcal U}} p(u,z) p(u|z) \sum_x p^2(x|uz) \\
    & \ge & \sum_{(z,u)\in A} p(u|z) p(u,z) \sum_x p^2(x|uz) \\
    & \ge &  \frac{2^{-t}}{|{\mathcal U}|} \sum_{(z,u)\in A} p(u,z)
    P_c(X|U=z,Z=z).
  \end{eqnarray*}
Thus
  \begin{equation} \label{pcbound}
    \sum_{(z,u)\in A} p(u,z) P_c(X|U=z,Z=z) \le
    2^{t}|{\mathcal U}|\avg{P_c(X|Z=z)}_z.
  \end{equation}
We can now use this result to bound $H(K|GUZ)$ as follows:
  \begin{eqnarray*}
    H(K|GUZ) & = & \sum_{u,z} p(u,z) H(K|G,U=u,Z=z)  \\
    & = & \sum_{(u,z)\in A}p(u,z) H(K|G,U=u,Z=z) + \sum_{(u,z)\in A^c}
    p(u,z) H(K|G,U=u,Z=z)\\
    & \ge & \sum_{(u,z)\in A} p(u,z) H(K|G,U=u,Z=z),
  \end{eqnarray*}
using the positivity of the conditional entropy functions, and the
fact that $U$ and $Z$ are independent of $G$. Plugging
(\ref{GUZbound}) into the above inequality leads to
  \begin{eqnarray*}
    H(K|GUZ) & \ge & \sum_{(u,z)\in A} p(u,z) \left(r - \frac{2^r}{\ln
    2}p_c(X|U=u,Z=z)\right) \\
    & \ge & (1-2^{-t})r - \frac{2^r}{\ln
    2}2^t|{\mathcal U}| \avg{p_c(X|Z=z)}_z \\
    & = & (1-2^{-t})r - 2^{r + t + \log_2 |{\mathcal U}| + \log_2
    \avg{p_c(X|Z=z)}_z},
  \end{eqnarray*}
as follows from (\ref{pcbound}).  We can then set
  \begin{equation}  \label{compression}
    r = -\log_2\avg{p_c(X|Z=z)}_z - t - \kappa - s,
  \end{equation}
where $\kappa=\log_2 |\mathcal{U}|$ is the number of bits in
message $U$ and $s$ is another security parameter.  This leads to
the bound
  \begin{equation}
    H(K|GUZ) \ge (1-2^{-t})r - \frac{2^{-s}}{\ln 2}.
  \end{equation}
Eve's mutual information can now be bounded by
  \begin{eqnarray*}
    I_E(K;GUZ) & = & H(K) - H(K|GUZ) \\
    & \le & 2^{-t}r + \frac{2^{-s}}{\ln 2}.
  \end{eqnarray*}
Plugging (~\ref{CPfactor}) into (\ref{compression}) leads directly
to
  \begin{equation}
    r = n\tau - t - \kappa - s,
  \end{equation}
where $\tau=-\log_2p_c$.

\section{Separating collision probability into photon number
contributions} \label{coherence}

We begin by first restating a theorem proven in the Appendix
of~\cite{Lutkenhaus96} which will play an important role in this
and some of the following appendices. This theorem states that the
expected collision probability can only increase in the presence
of more detailed knowledge. Mathematically it is stated as
follows: if the joint probability of signal $i$ and measurement
outcome $l$ is split up into two measurements $l'$ and $l"$ as
$p(i,l)=p(i,l')+p(i,l")$ for all $i$ then the average collision
probability can only increase. Multiple applications of this
theorem show that if a joint probability for all signals is broken
up into more measurements this can only improve the average
collision probability.

We now define the projectors
  \begin{eqnarray}
    E_{i}^{a} & = & \sum_{x=0}^i \|x,i-x>_a\<x,i-x| \\
    E_{i}^{b} & = & \sum_{y=0}^i \|y,i-y>_b\<y,i-y|,
  \end{eqnarray}
which projects modes $a$ and $b$ onto the the subspace where the
total photon number is $i$.  It is clear from the above definition
that
  \begin{equation}
    \sum_{i = 0}^{\infty} E_{i}^a = \sum_{i = 0}^{\infty} E_{i}^b =
    I.
  \end{equation}

Eve will generate a density matrix $\rho_{abe}$ which represents
the combined state of Alice and Bob's signal and Eve's measurement
probe.  We expanded $\rho_{abe}$ as
  \begin{equation}\label{RhoExpansion}
    \rho_{abe} = \sum_{i,j=0}^{\infty} \rho_{abe} E_{i}^a E_{j}^b.
  \end{equation}
We now prove that if $\rho_{abe}$ is replaced by the density
matrix
  \begin{equation} \label{rho2}
    \rho_{abe}' = \sum_{i,j=0}^{\infty} E_{i}^a E_{j}^b \rho_{abe}
    E_{i}^a E_{j}^b,
  \end{equation}
this has no effect on the measurements of Alice, Bob, or Eve. Note
that $\rho_{abe}'$ is the same density matrix as $\rho_{abe}$
minus the off diagonal terms which couple states in different
photon number spaces.

To prove our claim let $\psi_a$ and $\psi_b$ be the measurement
results of Alice and Bob. They could represent the reception of an
$x$, $y$, $u$, or $v$ polarized photon, or they could represent an
ambiguous dual fire detection. We define $F_{\psi_a}$ and
$F_{\psi_b}$ as the POM's corresponding to these different
measurement results. All the POM's which Alice can perform are of
the form
  \begin{equation}
    F_{\psi_a} = \sum_{m,n=0}^{\infty} \alpha_{mn}\|m,n>_+\<m,n|
    + \beta_{mn}\|m,n>_{\times}\<m,n|,
  \end{equation}
where $\|m,n>_+$ and $\|m,n>_\times$ represent number occupations
in the $x$-$y$ and $u$-$v$ basis respectively.  We can rewrite
this as
  \begin{eqnarray*}
    F_{\psi_a} & = & \sum_{i=0}^{\infty} E_i^a \sum_{m,n=0}^{\infty} \alpha_{mn}\|m,n>_+\<m,n|
    + \beta_{mn}\|m,n>_{\times}\<m,n| \\
    & = &  \sum_{i=0}^{\infty} E_i^a \sum_{m+n=i} \alpha_{mn}\|m,n>_+\<m,n|
    + \beta_{mn}\|m,n>_{\times}\<m,n| \\
    & = &  \sum_{i=0}^{\infty} E_i^a \left( \sum_{m+n=i} \alpha_{mn}\|m,n>_+\<m,n|
    + \beta_{mn}\|m,n>_{\times}\<m,n|\right)E_i^a \\
    & = & \sum_{i=0}^{\infty} E_i^a F_i^a E_i^a.
  \end{eqnarray*}
Similarly we can write
  \begin{equation}
    F_{\psi_b} = \sum_{i=0}^{\infty} E_i^b F_i^b E_i^b.
  \end{equation}
The joint probability that Alice measures result $\psi_a$, Bob
measures $\psi_b$, and Eve finds her probe in state $k$ is
  \begin{equation}
    p(\psi_a, \psi_b, k) = \TR{\rho_{abe} F_{\psi_a}
    F_{\psi_b} E_k},
  \end{equation}
where $E_k$ is the projection onto state $k$ of Eve's probe.  This
can be rewritten as
  \begin{eqnarray*}
    p(\psi_a, \psi_b, k) & = & \sum_{i,j} \TR{\rho_{abe} E_i^a E_j^b F_{\psi_a}
    F_{\psi_b} E_k} \\
    & = & \sum_{i,j} \TR{\rho_{abe} E_i^a E_j^b \sum_{mn} E_m^a E_n^b F_m^a F_n^b E_m^a E_n^b
    E_k}\\
    & = & \sum_{i,j} \TR{\rho_{abe} E_i^a E_j^b F_i^a F_j^b E_k E_i^a E_j^b }
    \\
    & = & \sum_{i,j} \TR{E_i^a E_j^b \rho_{abe} E_i^a E_j^b F_i^a F_j^b E_k
    }\\
    & = & \sum_{i,j} \TR{E_i^a E_j^b \rho_{abe} E_i^a E_j^b F_{\psi_a} F_{\psi_b} E_k
    }\\
    & = & \TR{\rho_{abe}' F_{\psi_a} F_{\psi_a} E_k}.
  \end{eqnarray*}
Thus, both matrices result in the same joint probability.  This
means that the most general density matrix can be written in block
diagonal form
  \begin{eqnarray*}
    \rho_{abe} & = & \sum_{i,j=1}^{\infty} E_{i}^a E_{j}^b \rho_{abe}
    E_{i}^a E_{j}^b \\
    & = & \sum_{i,j=1}^{\infty} \rho_{abe}^{(ij)},
  \end{eqnarray*}
where the matrix $\rho_{abe}^{(ij)}$ is a matrix over the entire
hibert space of Eve's probe, and the subspace in which Alice
receives $i$ photons and bob receives $j$ photons.

We now show that Eve's optimal strategy will be to keep track of
how many photons she is sending to Alice and Bob.  We define
$\|\mu_k>$ as the measurement basis over which Eve will measure
her probe, and $F_\psi$ as the positive operator measurement
performed by Alice's detection unit. We then have, using the above
expansion
  \begin{eqnarray*}
    p(\psi,\mu_k) & = & \TR{ \rho_{abe}F_{\psi}\kb|\mu_k><\mu_k| }
    \\
    & = & \sum_{ij} \TR{\rho_{abe}^{(ij)} F_{\psi} \kb|\mu_k><\mu_k|} \\
    & = & \sum_{ij} \TR{ \<\mu_k|\rho_{abe}^{(ij)}\|\mu_k> F_{\psi}
    }.
  \end{eqnarray*}
Eve can always construct a new state which will perform at least
as well as the above and will allow her to keep track of how many
photons she sent. Define the new density matrix
  \begin{equation}
    \tilde{\rho}_{abe} = \sum_{ij} \tilde{\rho}_{abe}^{(ij)},
  \end{equation}
where
  \begin{equation}
    \tilde{\rho}_{abe}^{(ij)} =
    \rho_{abe}^{(ij)}\otimes\aux|ij><ij|,
  \end{equation}
and the term $\aux|ij><ij|$ is the state of an auxiliary system
which keeps track of the number of photons sent.  This new state
will not change the measurement outcome of Alice or Bob because
tracing out over the auxiliary system leads to the original
density matrix.  We define the new measurement basis as
  \begin{equation}
    \|\mu_k^{(mn)}> = \|\mu_k>\otimes\|mn>_{a}.
  \end{equation}
If Eve measures in this new basis we have
  \begin{equation}
    p(\psi, \mu_k^{(mn)}) = \TR{\<\mu_k^{(mn)}|
    \tilde{\rho}_{abe} \|\mu_k^{(mn)}> F_{\psi}},
  \end{equation}
and using the above definitions it is easy to show that
  \begin{equation}
    \<\mu_k^{(mn)}| \tilde{\rho}_{abe} \|\mu_k^{(mn})> = \<\mu_k|
    \rho_{abe}^{(ij)} \|\mu_k> \delta_{im} \delta_{jn}.
  \end{equation}
Combining these two equations one gets
  \begin{equation}
    p(\psi,\mu_k^{(mn)}) = \TR{\<\mu_k|\rho_{abe}^{(mn)} \|\mu_k>
    F_{\psi}}.
  \end{equation}
Furthermore,
  \begin{equation}
    p(\psi,\mu_k) = \sum_{mn} p(\psi,\mu_k^{(mn)}).
  \end{equation}
The measurement $\mu_k$ has been split up into more detailed
measurements whose probabilities add up to the original.  As
stated earlier this more detailed information can only lead to an
increase in the average collision probability, thus the new probe
and measurement basis must be at least as good as the old one.

\section{Bound on one photon contribution for collision
probability} \label{OnePhoton}

In this Appendix we put a bound on $p_c^{(11)}$ which is defined
as the contribution to the collision probability from signals in
which Alice and Bob each receive one photon.  We assume that Eve
can store her probe coherently until until she learns all relevant
information from public discussion. Her only restriction is that
she must measure each probe independently. During the public
discussion, Eve will learn the measurement basis used by Alice and
Bob. She will also learn which bits were received correctly, and
which incorrectly from the error correction phase. This
information can potentially refine her measurement by allowing her
to split bits into groups which will receive different treatment.
A different measurement basis will be used for each case. We will
define this basis as
  \begin{eqnarray*}
    \|a_k> & - & \mbox{bit received correctly in x-y basis}\\
    \|b_k> & - & \mbox{bit received correctly in u-v basis}\\
    \|c_k> & - & \mbox{bit received incorrectly in x-y
    basis} \\
    \|d_k> & - & \mbox{bit received incorrectly in u-v
    basis}
  \end{eqnarray*}
Each signal sent has a probability $p_{rec}^{(11)}$ of entering
the reconciled (error corrected) key. The signal can enter the key
as a correct transmission or an error which will happen with
probability $p_{err}^{(11)}$.

We first assume that $\rho_{abe}^{(11)}$ is a pure state.  There
is no loss of generality in this because for any mixed state one
can construct a pure state which is at least as good.  We can show
this by first expanding $\rho_{abe}^{(11)}$ in pure states
  \begin{equation}
    \rho_{abe}^{(11)} = \sum_i \sigma_i \kb|\psi_i><\psi_i|.
  \end{equation}
If Eve uses the measurement basis $\|\mu_k>$ then
  \begin{equation}
    p(\psi,\mu_k) = \sum_i \sigma_i
    \TR{\bk<\mu_k|\psi_i>\bk<\psi_i|\mu_k> F_\psi}.
  \end{equation}
Suppose now that instead of sending the mixed state
$\rho_{abe}^{(11)}$ Eve sends the pure state
  \begin{equation}
    \|\psi>_{abe} = \sum_i \sqrt{\sigma_i} \|\psi_i>\otimes\|i>,
  \end{equation}
where $\|i>$ are the eigenstates of an additional system which
keeps track of which pure state was sent.  One can define a new
measurement basis
  \begin{equation}
    \|\mu_k^i> = \|\mu_k>\otimes\|i>.
  \end{equation}
It is easy to show that
  \begin{equation}
    p(\psi,\mu_k) = \sum_i p(\psi,\mu_k^i).
  \end{equation}
It is also clear that this new state does not change the
measurement outcomes for Alice and Bob, because tracing out the
additional system results in the same density matrix as before.
Thus, the new state must be at least as good.

Rather than using the more cumbersome occupation number notation,
we will adopt a shorthand notation for the case where only one
photon is sent to either Alice or Bob. We will use the eigenstates
$\|xx>$, $\|yy>$, $\|xy>$, $\|yx>$ to denote the different
polarization states.  Starting with the most generic state
$\|\psi>_{abe}$, we expand it in the polarization basis as
  \begin{eqnarray} \label{Rho11}
    \|\psi> & = & \|xx>\bk<xx|\psi> + \|yy>\bk<yy|\psi> +
    \|xy>\bk<xy|\psi> + \|yx>\bk<yx|\psi> \\
    & = & \|xx>\|P_{xx}> + \|yy>\|P_{yy}> + \|xy>\|P_{xy}> +
    \|yx>\|P_{yx}>,
  \end{eqnarray}
where the states $\|P_{st}>$ represent Eve's probe and are not
necessarily normalized or orthogonal.  It should be noted that the
above state is not normalized to $1$ but
  \begin{equation}
    \bk<P_{xx}|P_{xx}> + \bk<P_{yy}|P_{yy}> + \bk<P_{xy}|P_{xy}> +
    \bk<P_{yx}|P_{yx}> = p_{11},
  \end{equation}
where $p_{11}$ is the probability that Eve only sends one photon
to Alice and Bob.   Using the relationships
  \begin{eqnarray}
    \|u> & = & \frac{1}{\sqrt{2}} \left( \|x> + \|y> \right) \\
    \|v> & = & \frac{1}{\sqrt{2}} \left( \|x> - \|y> \right),
  \end{eqnarray}
one can rewrite (\ref{Rho11}) as
  \begin{equation}
    \|\psi> = \|uu>\|P_{uu}> + \|vv>\|P_{vv}> + \|uv>\|P_{uv}> +
    \|vu>\|P_{vu}>,
  \end{equation}
where
  \begin{eqnarray}
    \|P_{uu}> & = & \frac{1}{2} \left( \|P_{xx}> + \|P_{yy}> + \|P_{xy}> +
    \|P_{yx}> \right) \label{xy2uv1} \\
    \|P_{vv}> & = & \frac{1}{2} \left( \|P_{xx}> + \|P_{yy}> - \|P_{xy}> -
    \|P_{yx}> \right) \label{xy2uv2} \\
    \|P_{uv}> & = & \frac{1}{2} \left( \|P_{xx}> - \|P_{yy}> - \|P_{xy}> +
    \|P_{yx}> \right) \label{xy2uv3} \\
    \|P_{vu}> & = & \frac{1}{2} \left( \|P_{xx}> - \|P_{yy}> + \|P_{xy}> -
    \|P_{yx}> \right).  \label{xy2uv4}
  \end{eqnarray}

We will introduce the notation
  \begin{equation}
    \bk<a_k|P_{xx}>  = P_{xx}^k,
  \end{equation}
and use the same notation for all other projections.   Without
loss of generality we can assume that the projections $P_{st}^k$
are real. If these projection are complex numbers then we can
always find a probe of higher dimensionality which has real
projections and is at least as good, using the same trick of
probability splits described in Appendix~\ref{coherence}. Take for
example the projection $P_{xx}^k$.  We can rewrite this as
  \begin{equation} \label{ReIm}
    p(x,a_k) = \frac{1}{4}\Re[P_{xx}^k]^2 +
    \frac{1}{4}\Im[P_{xx}^k]^2.
  \end{equation}
We expand $\|a_k>$ and $\|P_{xx}>$ in the some orthogonal basis
$\|v_i>$ so that
  \begin{eqnarray*}
    \|P_{xx}>  & = & \sum_i \alpha_i \|v_i> \\
    \|a_k> & = & \sum_i \beta_i \|v_i>,
  \end{eqnarray*}
and define
  \begin{eqnarray*}
    \|P_{xx}^*>  & = & \sum_i \alpha_i^* \|v_i> \\
    \|a_k^*> & = & \sum_i \beta_i^* \|v_i>.
  \end{eqnarray*}
We can then define a new probe
  \begin{equation}
    \|\tilde{P}_{xx}> = \frac{1}{\sqrt{2}} \left(
    \|P_{xx}>\otimes\|x> + \|P_{xx}^*>\otimes\|y> \right),
  \end{equation}
and two new measurements
  \begin{eqnarray}
    \|a_k^{\Re} > & = & \frac{1}{\sqrt{2}} \left(
    \|a_k>\otimes\|x> +  \|a_k^*>\otimes\|y>
    \right) \\
    \|a_k^{\Im} > & = & \frac{1}{\sqrt{2}} \left(
    \|a_k>\otimes\|x> -  \|a_k^*>\otimes\|y>
    \right).
  \end{eqnarray}
It is easy to show, using (\ref{ReIm}), that the projections on
this new probe are all real and that
  \begin{equation}
    p(x,a_k) = p(x,a_k^{\Re}) + p(x,a_k^{\Im}).
  \end{equation}
This state must be at least as good for the aforementioned reason,
so an optimal solution exists which has only real projections. We
can now write all the relevant probabilities as
  \begin{eqnarray}
    p(x,a_k) &=& \frac{1}{4} \left(P_{xx}^k\right)^2 \\
    p(y,a_k) &=& \frac{1}{4} \left(P_{yy}^k\right)^2 \\
    p(x,c_k) &=& \frac{1}{4} \left(P_{xy}^k\right)^2 \\
    p(y,c_k) &=& \frac{1}{4} \left(P_{yx}^k\right)^2.
  \end{eqnarray}
The probabilities in the $u$-$v$ basis are obtained by replacing
$x$ and $y$ with $u$ and $v$ and $a$ and $c$ with $b$ and $d$
respectively.  These define all the probabilities which will
factor into $p_c^{(11)}$.

We can now write the collision probability in terms of the above
expressions
  \begin{equation}
    p_c^{(11)} = \frac{1}{4p_{rec}^{(11)}} \sum_k \left(
    \frac{\left(P_{xx}^k\right)^4 +\left(P_{yy}^k\right)^4}
    {\left(P_{xx}^k\right)^2 + \left(P_{yy}^k\right)^2} +
    \frac{\left(P_{uu}^k\right)^4 +\left(P_{vv}^k\right)^4}
    {\left(P_{uu}^k\right)^2 + \left(P_{vv}^k\right)^2} +
    \frac{\left(P_{xy}^k\right)^4 +\left(P_{yx}^k\right)^4}
    {\left(P_{xy}^k\right)^2 + \left(P_{yx}^k\right)^2} +
    \frac{\left(P_{uv}^k\right)^4 +\left(P_{vu}^k\right)^4}
    {\left(P_{uv}^k\right)^2 + \left(P_{vu}^k\right)^2} \right).
  \end{equation}
Using the Cauchy inequality discussed
in~\cite[Appendix]{Lutkenhaus99} we can put an upper bound on the
above expression of the form
  \begin{eqnarray*}
    p_c^{(11)} & \le & 1 - \frac{1}{2p_{rec}^{(11)}} \left(
    \frac{\left(\sum_k P_{xx}^k P_{yy}^k\right)^2}{\sum_k
    \left(P_{xx}^k\right)^2 + \sum_k \left(P_{xx}^k\right)^2} +
    \frac{\left(\sum_k P_{uu}^k P_{vv}^k\right)^2}{\sum_k
    \left(P_{uu}^k\right)^2 + \sum_k \left(P_{vv}^k\right)^2} \right. + \\
    & & \left.
    \frac{\left(\sum_k P_{xy}^k P_{yx}^k\right)^2}{\sum_k
    \left(P_{xy}^k\right)^2 + \sum_k \left(P_{yx}^k\right)^2} +
    \frac{\left(\sum_k P_{uv}^k P_{vu}^k\right)^2}{\sum_k
    \left(P_{uv}^k\right)^2 + \sum_k \left(P_{vu}^k\right)^2}
    \right).
  \end{eqnarray*}
However we notice that
  \begin{equation}
    \sum_k P_{st}^k P_{s't'}^k = \sum_k \bk<P_{st}|\mu_k^{(11)}>
    \bk<\mu_k^{(11)}|P_{s't'}> = \bk<P_{st}|P_{s't'}>,
  \end{equation}
using the completeness of Eve's measurement basis.  Thus our new
bound on the collision probability is
  \begin{eqnarray*}
    p_c^{(11)} & \le & 1 - \frac{1}{2p_{rec}^{(11)}} \left(
    \frac{\bk<P_{xx}|P_{yy}>^2}{\bk<P_{xx}|P_{xx}> + \bk<P_{yy}|P_{yy}>} +
    \frac{\bk<P_{uu}|P_{uu}>^2}{\bk<P_{uu}|P_{uu}> + \bk<P_{vv}|P_{vv}>}
    + \right. \\
    & & \left.
    \frac{\bk<P_{xy}|P_{yx}>^2}{\bk<P_{xy}|P_{xy}> + \bk<P_{yx}|P_{yx}>} +
    \frac{\bk<P_{uv}|P_{vu}>^2}{\bk<P_{uv}|P_{uv}> + \bk<P_{vu}|P_{vu}>}
    \right).
  \end{eqnarray*}
The relations in (\ref{xy2uv1})-(\ref{xy2uv4}) can be used to
replace the $u$-$v$ terms with $x$-$y$ terms.  We also impose the
symmetric eavesdropping conditions
  \begin{eqnarray}
    \bk<P_{xx}|P_{xx}> & = & \bk<P_{yy}|P_{yy}> \\
    \bk<P_{xy}|P_{xy}> & = & \bk<P_{yx}|P_{yx}>,
  \end{eqnarray}
which state that Eve must keep the number of $x$ and $y$ states
balanced.  If Eve does not maintain these conditions, she will be
immediately detected because Alice and Bob will note an asymmetry
in their measurements.  It is also shown in~\cite{Lutkenhaus99}
that these are not restrictions because any state which does not
satisfy the above symmetry conditions can be replaced by one that
does and which is at least as good.  Taking into account all these
conditions one sees that there are very few degrees of freedom
left to optimize.  They are $\bk<P_{xx}|P_{xx}>$,
$\bk<P_{xy}|P_{xy}>$, and the angles $\varphi_{xx}^{yy}$, and
$\varphi_{xy}^{yx}$.  The probability that a signal will enter the
reconciled key is
  \begin{equation}
    p_{rec}^{(11)} = \frac{1}{2} \left(\bk<P_{xx}|P_{xx}> +
    \bk<P_{yy}|P_{yy}> + \bk<P_{xy}|P_{xy}> + \bk<P_{yx}|P_{yx}>
    \right),
  \end{equation}
and the probability that it will enter the sifted key as an error
is given by
  \begin{equation}
    p_{err}^{(11)} = p_{rec}^{11} - \frac{1}{4}\left(\bk<P_{xx}|P_{xx}> +
    \bk<P_{yy}|P_{yy}> + \bk<P_{uu}|P_{uu}> +
    \bk<P_{vv}|P_{vv}> \right).
  \end{equation}
The above relations can be directly plugged into the definition of
the disturbance to give
  \begin{equation} \label{eps11}
    \epsilon^{(11)} = \frac{\bk<P_{xx}|P_{xx}>(1-\cos\varphi_{xx}^{yy}) +
    \bk<P_{xy}|P_{yx}>(3-\cos\varphi_{xy}^{yx}) }{4 \left(
    \bk<P_{xx}|P_{xx}>+\bk<P_{xy}|P_{yx}> \right)}.
  \end{equation}
The collision probability is then bounded by
  \begin{eqnarray*}
    p_c^{(11)} & \le & \frac{3}{4} -
    \frac{\bk<P_{xx}|P_{xx}>\cos^2\varphi_{xx}^{yy} +
    \bk<P_{xy}|P_{yx}>\cos^2\varphi_{xy}^{yy}}{4 \left(
    \bk<P_{xx}|P_{xx}>+\bk<P_{xy}|P_{yx}> \right)} + \\
    & & \frac{\bk<P_{xx}|P_{xx}>\bk<P_{xy}|P_{yx}>}{
    2\left(\bk<P_{xx}|P_{xx}>+\bk<P_{xy}|P_{yx}>\right)} \left[
    \frac{(1+\cos \varphi_{xx}^{yy})(1+\cos\varphi_{xy}^{yx})}
    {\bk<P_{xx}|P_{xx}>(1+\cos\varphi_{xx}^{yy}) +
    \bk<P_{xy}|P_{yx}>(1+\cos\varphi_{xy}^{yx}) }\right. \\
    & & \left. +
    \frac{(1-\cos \varphi_{xx}^{yy})(1-\cos\varphi_{xy}^{yx})}
    {\bk<P_{xx}|P_{xx}>(1-\cos\varphi_{xx}^{yy}) +
    \bk<P_{xy}|P_{yx}>(1-\cos\varphi_{xy}^{yx}) }
    \right].
  \end{eqnarray*}
The right hand side of the above equation should be maximized
subject to the constraint given in Equation~\ref{eps11}.  As shown
in~\cite{Lutkenhaus99}, the maximum is achieved when
$\cos\varphi_{xx}^{yy}=\cos\varphi_{xy}^{yx}=1-2\epsilon^{11}$ and
$\bk<P_{xx}|P_{xx}>=\bk<P_xy|P_xy>(1-\epsilon^{11})/\epsilon^{11}$.
The resulting bound on the collision probability is
  \begin{equation}
    p_c^{(11)} \le \frac{1}{2} + 2\epsilon^{(11)} -
    2(\epsilon^{(11)})^2.
  \end{equation}
The above equation indicates that for $\epsilon^{(11)}=1/2$ Eve
can have complete knowledge over Alice's key.  This can be
accomplished by sending Alice one of a pair of maximally entangled
photons and keeping the other, while sending Bob a third photon
with completely random polarization.  After the measurement basis
is revealed, a measurement of the retained photon will tell the
bit value of Alice's string.

\section{Higher order number state
contributions}\label{MultiPhoton}

Higher number states are taken into account by setting $w_D$
sufficiently large so that Eve's optimal strategy is to only use
single photon states.  If Eve sends $n$ photons to Alice or Bob,
the probability that all $n$ photons will be measured in the same
basis is $2\times2^{-n}$.  If Eve sends $i$ photons to Alice and
$j$ photons to Bob we have
  \begin{eqnarray}
    p_D^{ij} & \ge & 2\left(\frac{1}{2} -
    \left(\frac{1}{2}\right)^i\right)\left(\frac{1}{2} -
    \left(\frac{1}{2}\right)^j\right)\TR{\rho_{abe}^{ij}} \\
    p_{rec}^{ij} & \le &
    2\left(\frac{1}{2}\right)^i\left(\frac{1}{2}\right)^j
    \TR{\rho_{abe}^{ij}},
  \end{eqnarray}
which leads to
  \begin{equation}
    \frac{p_D^{ij}}{p_{rec}^{ij}}
    \ge  \frac{ \left(\frac{1}{2} - \left(\frac{1}{2}
    \right)^i\right)}{\left(\frac{1}{2}\right)^i}
    \frac{ \left(\frac{1}{2} - \left(\frac{1}{2}\right)^j\right)}{
    \left(\frac{1}{2}\right)^j} \ge 1.
  \end{equation}
As noted above a disturbance of $1/2$ already implies that Eve can
obtain the entire string.  So setting $w_D$ to $1/2$ means that
Eve can do at least as good by sending only one photon to Alice of
Bob.  Thus
  \begin{equation}
    p_c \le \frac{1}{2} + 2\epsilon - 2\epsilon^2.
  \end{equation}

\end{appendix}

\bibliography{references}

\begin{thebibliography}{10}

\bibitem{Bennett92}
C.H. Bennett.
\newblock Quantum cryptography using any two nonorthogonal states.
\newblock {\em Phys. Rev. Lett.}, 68(21):3121--3124, May 1992.

\bibitem{BennettBrassard84}
C.H. Bennett and G.~Brassard.
\newblock {\it Proceedings of IEEE International Conference on Computers,
  Systems, and Signal Processing, Bangalore, India (IEEE, New York, 1984)}, pp.
  175-179, 1984.

\bibitem{BennettBrassard95}
C.H. Bennett, G.~Brassard, C.~Crépeau, and U.M. Maurer.
\newblock Generalized privacy amplification.
\newblock {\em IEEE Trans. Inf. Theory}, 41(6):1915--1923, 1995.

\bibitem{BennettBrassard92}
C.H. Bennett, G.~Brassard, and N.D. Mermin.
\newblock Quantum cryptography without bell's theorem.
\newblock {\em Phys. Rev. Lett.}, 68(5):557--559, 1992.

\bibitem{BensonSantori00}
O.~Benson, C.~Santori, M.~Pelton, and Y.~Yamamoto.
\newblock Regulated and entangled photons from a single quantum dot.
\newblock {\em Phys. Rev. Lett.}, 84(11):2513--2516, March 2000.

\bibitem{BihamBoyer00}
E.~Biham, M.~Boyer, P.O. Boykin, T.~Mor, and V.~Roychowder.
\newblock {\it Proceedings of the Thirty-Second Annual ACS Symposium of Theory
  of Computing (ACM Press, New York, 2000)}, pp. 715-724.

\bibitem{BihamMor97}
E.~Biham and T.~Mor.
\newblock Security of quantum cryptography against collective attacks.
\newblock {\em Phys. Rev. Lett.}, 78(11):2256--2259, March 1997.

\bibitem{BourennaneGibson99}
M.~Bourennane, F.~Gibson, A.~Karlsson, A.~Hening, P.~Johnson, T.~Tsegaye,
  D.~Ljunggren, and E.~Sundberg.
\newblock {\em Opt. Express}, 4:383, 1999.

\bibitem{BrassardLutkenhaus00}
G.~Brassard, N.~L{\"{u}}tkenhaus, T.~Mor, and B.C. Sanders.
\newblock Limitations on practical quantum cryptography.
\newblock {\em Phys. Rev. Lett.}, 85(6):1330--1333, August 2000.

\bibitem{BrassardSalvail93}
G.~Brassard and L.~Salvail.
\newblock Secret-key reconciliation by public discussion.
\newblock Advances in Cryptology -- EUROCRYPT '93, Vol. 765 of Lecture Notes in
  Computer Science, edited by T. Hellseth (Springer, Berlin, 1994), pp.
  410-423, 1993.

\bibitem{BriegelDur98}
H.~J. Briegel, W.~D{\"{u}}r, J.I. Cirac, and P.~Zoller.
\newblock Quantum repeaters: The role of imperfect local operations in quantum
  communication.
\newblock {\em Phys. Rev. Lett.}, 81(26):5932--5935, December 1998.

\bibitem{ButtlerHughes00}
W.T. Buttler, R.J. Hughes, S.K. Lamoreaux, G.L. Morgan, J.E. Nordholt, and C.G.
  Peterson.
\newblock Daylight quantum key distribution over 1.6km.
\newblock {\em Phys. Rev. Lett.}, 84(24):5652--5655, June 2000.

\bibitem{Cachin97}
C.~Cachin and U.M. Maurer.
\newblock Linking information reconciliation and privacy amplification.
\newblock {\em J. Cryptology}, 10:97, 1997.

\bibitem{CalsamigliaLutkenhaus00}
J.~Calsamiglia and N.~Lutkenhaus.
\newblock Maximum efficiency of a linear-optical bell-state analyzer.
\newblock e-print quant-ph/0007058, July 2000.

\bibitem{Cover}
T.M. Cover.
\newblock {\em Elements of information theory}.
\newblock Wiley, New York, 1991.

\bibitem{Ekert91}
A.K. Ekert.
\newblock Quantum cryptography based on bell's theorem.
\newblock {\em Phy. Rev. Lett.}, 67(6):661--663, August 1991.

\bibitem{Ekert94}
A.K. Ekert, B.~Huttner, G.M. Palma, and A.~Peres.
\newblock Eavesdropping on euantum-cryptographical systems.
\newblock {\em Phys. Rev. A}, 50(2):1047--1056, 1994.

\bibitem{Fuchs97}
C.A. Fuchs, N.~Gisin, R.B. Griffiths, C.S. Niu, and A.~Peres.
\newblock Optimal eavesdropping in quantum cryptography. i. information bound
  and optimal strategy.
\newblock {\em Phys. Rev. A}, 56(2):1163--1172, 1997.

\bibitem{Helstrom76}
C.W. Helstrom.
\newblock Operator-valued measures in quantum decision and estimation theory.
\newblock {\em Notices of the American Mathematical Society}, 23(1):A163, 1976.

\bibitem{HuttnerEkert94}
B.~Huttner and A.K. Ekert.
\newblock Information gain in quantum eavesdropping.
\newblock {\em J. Mod. Optics}, 41(12):2455--2466, 1994.

\bibitem{HuttnerImoto95}
B.~Huttner, N.~Imoto, N.~Gisin, and T.~Mor.
\newblock Quantum cryptography with coherent states.
\newblock {\em Phy. Rev. A}, 51(3):1863--1869, March 1995.

\bibitem{JenneweinSimon00}
T.~Jennewein, C.~Simon, G.~Weihs, H.~Weinfurter, and A.~Zeilinger.
\newblock Quantum cryptography with entangled photons.
\newblock {\em Phys. Rev. Lett.}, 84(20):4729--4732, May 2000.

\bibitem{KwiatMattle95}
P.G. Kwiat, K.~Mattle, H.~Weinfurter, and A.~Zeilinger.
\newblock New high-intensity source of polarization entangled photon pairs.
\newblock {\em Phys. Rev. Lett.}, 75(24):4337--4341, December 1995.

\bibitem{LoChau99}
H.K. Lo and H.F. Chau.
\newblock Unconditional security of quantum key distribution over arbitrarily
  long distances.
\newblock {\em Science}, 283:2050--2056, March 1999.

\bibitem{Lutkenhaus96}
N.~L{\"{u}}tkenhaus.
\newblock Security against eavesdropping in quantum cryptography.
\newblock {\em Phys. Rev. A}, 54(1):97--111, 1996.

\bibitem{Lutkenhaus99}
N.~L{\"{u}}tkenhaus.
\newblock Estimates for practical quantum cryptography.
\newblock {\em Phys. Rev. A}, 59(5):3301--3319, 1999.

\bibitem{Lutkenhaus00}
N.~L{\"{u}}tkenhaus.
\newblock Security against individual attacks for realistic quantum key
  distribution.
\newblock {\em Phys. Rev. A}, 61(5):2304--, 2000.

\bibitem{MarandTownsend95}
C.~Marand and P.T. Townsend.
\newblock {\em Opt. Lett.}, 20:1695, 1995.

\bibitem{Mayers96}
D.~Mayers.
\newblock e-print quant-ph/9802025.

\bibitem{MayersYao98}
D.~Mayers and A.~Yao.
\newblock Quantum cryptography with imperfect apparatus.
\newblock e-print quant-ph/0007058, September 1998.

\bibitem{NaikPeterson00}
D.S. Naik, C.G. Peterson, A.G. White~A.J. Berglung, and P.G. Kwiat.
\newblock Entangled state quantum cryptography: Eavesdropping on the ekert
  protocol.
\newblock {\em Phys. Rev. Lett.}, 84(20):4733--4736, May 2000.

\bibitem{PanBouwmeester98}
J.W. Pan, D.~Bouwmeester, H.~Weifurter, and A.~Zeilinger.
\newblock Experimental entanglement swapping: Entangling photons that never
  interacted.
\newblock {\em Phys. Rev. Lett.}, 80(18):3891--3894, May 1998.

\bibitem{RigbordyGautier00}
G.~Rigbordy, J.D. Gautier, N.~Gisin, O.~Guinnard, and H.~Zbinden.
\newblock {\em J. Mod. Opt.}, 47:517, 2000.

\bibitem{ShorPreskill00}
P.W. Shor and J.~Preskill.
\newblock Simple proof of security of the bb84 quantum key distribution
  protocol.
\newblock {\em Phys. Rev. Lett.}, 85(2):441--444, July 2000.

\bibitem{Slutsky98}
B.A. Slutsky, R.~Rao, P.C. Sun, and Y.~Fainman.
\newblock Security of quantum cryptography against individual attacks.
\newblock {\em Phys. Rev. A}, 57(4):2383--2398, 1998.

\bibitem{TittelBrendel99}
W.~Tittel, J.~Brendel, N.~Gisin, and H.~Zbinden.
\newblock Long-distance bell-type tests using energy-time entangled photons.
\newblock {\em Phys. Rev. A}, 59(6):4150--4163, June 1999.

\bibitem{TittelBrendel00}
W.~Tittel, J.~Brendel, H.~Zbinden, and N.~Gisin.
\newblock Quantum cryptography using entangled photons in energy-time bell
  states.
\newblock {\em Phys. Rev. Lett.}, 84(20):4737--4740, May 2000.

\bibitem{Townsend98}
P.D. Townsend.
\newblock {\em IEE Photonics Technol. Lett.}, 10:1048, 1998.

\bibitem{WallsMilburn}
D.F. Walls and G.J. Milburn.
\newblock {\em Quantum Optics}.
\newblock Springer, Berlin, 1994.

\end{thebibliography}
\bibliographystyle{plain}

\end{document}